\documentclass[twocolumn]{aastex62}

\usepackage{graphicx}
\usepackage{amsmath, amssymb,amstext}
\usepackage{subfigure}
\usepackage{color}
\usepackage{xspace}
\usepackage{url}
\usepackage{hyperref}
\usepackage{array}
\usepackage{float}
\usepackage{threeparttable}
\usepackage{longtable}
\usepackage{nicefrac}
\usepackage{afterpage}
\usepackage{todonotes}

\newcommand{\angstrom}{\textup{\AA}\xspace}
\newcommand{\io}[2]{#1\,{\textsc{#2}}}

\def\kms{{\rm km\,s$^{-1}$}\xspace}
\def\ergs{{\rm\,erg\,s$^{-1}$}\xspace}

\def\Rsun{{\rm R$_{\odot}$}\xspace}
\def\Msun{{\rm M$_{\odot}$}\xspace}

\def\halpha{{\rm H$\alpha$}\xspace}
\def\he2{{He~{\small II}}\xspace}

\received{September 20, 2018}
\revised{\today}
\accepted{February 1, 2019}

\submitjournal{ApJ}

\shorttitle{Tidal Disruption Event \lowercase{i}PTF15\lowercase{af}}
\shortauthors{Blagorodnova et al.}

\begin{document}

\title{The Broad Absorption Line Tidal Disruption Event \lowercase{i}PTF15\lowercase{af}: Optical and Ultraviolet Evolution}

\correspondingauthor{Nadejda Blagorodnova}
\email{N.Blagorodnova@astro.ru.nl}

\author[0000-0003-0901-1606]{N. Blagorodnova}
 \altaffiliation{VENI Fellow}
 \affiliation{  Cahill Center for Astrophysics, California Institute of Technology, Pasadena, CA 91125, USA }
 \affiliation{Department of Astrophysics/IMAPP, Radboud University, Nijmegen, The Netherlands}

\author{S.~B.~Cenko}
\affiliation{Joint Space-Science Institute, University of Maryland, College Park, MD 20742, USA }
\affiliation{NASA Goddard Space Flight Center, Mail Code 661, Greenbelt, MD 20771, USA}

\author{ S.~R.~Kulkarni }
 \affiliation{  Cahill Center for Astrophysics, California Institute of Technology, Pasadena, CA 91125, USA }

\author{I.~Arcavi}
 \altaffiliation{Einstein Fellow}
\affiliation{Department of Physics, University of California, Santa Barbara, CA 93106-9530, USA}
 \affiliation{Las Cumbres Observatory, 6740 Cortona Dr Ste 102, Goleta, CA 93117-5575, USA}
 \affiliation{The School of Physics and Astronomy, Tel Aviv University, Tel Aviv 69978, Israel}

\author{J.~S.~Bloom}
\affiliation{Department of Astronomy, University of California, Berkeley, CA 94720-3411, USA}

\author[0000-0002-9256-6735]{G.~Duggan}
 \affiliation{  Cahill Center for Astrophysics, California Institute of Technology, Pasadena, CA 91125, USA }

 \author{A.~V.~Filippenko}
 \affiliation{Department of Astronomy, University of California, Berkeley, CA 94720-3411, USA}
 \affiliation{Miller Senior Fellow, Miller Institute for Basic Research in Science, University of California, Berkeley, CA  94720, USA}

 \author{C.~Fremling}
 \affiliation{  Cahill Center for Astrophysics, California Institute of Technology, Pasadena, CA 91125, USA }

 \author{A.~Horesh}
 \affiliation{Racah Institute of Physics, The Hebrew University of Jerusalem, Jerusalem, 91904, Israel}

\author[0000-0002-0832-2974]{G.~Hosseinzadeh}
 \affiliation{Harvard-Smithsonian Center for Astrophysics, 60 Garden Street, Cambridge, MA 02138-1516, USA}

 \author{E.~Karamehmetoglu}
 \affiliation{Department of Astronomy, The Oskar Klein Centre, Stockholm University, AlbaNova, 10691, Stockholm, Sweden}
 
\author{A.~Levan}
 \affiliation{Department of Physics, University of Warwick, Coventry, CV4 7AL, UK}

\author{F.~J.~Masci}
\affiliation{Infrared Processing and Analysis Center, California Institute of Technology, Pasadena, CA 91125, USA}

\author[0000-0002-3389-0586]{P.~E.~Nugent}
\affiliation{Computational Science Department, Lawrence Berkeley National Laboratory, 1 Cyclotron Road, MS 50B-4206, Berkeley, CA 94720, USA}
\affiliation{Department of Astronomy, University of California, Berkeley, CA 94720-3411, USA}

  \author{D.~R.~Pasham}
   \altaffiliation{Einstein Fellow}
 \affiliation{Massachusetts Institute of Technology, Cambridge, MA 02139, USA}

 \author{S.~Veilleux}
 \affiliation{Joint Space-Science Institute, University of Maryland, College Park, MD 20742, USA }
\affiliation{ Department of Astronomy, University of Maryland, Stadium Drive, College Park, MD 20742-2421, USA }

\author{R.~Walters}
 \affiliation{  Cahill Center for Astrophysics, California Institute of Technology, Pasadena, CA 91125, USA }

\author{ L.~Yan }
 \affiliation{  Cahill Center for Astrophysics, California Institute of Technology, Pasadena, CA 91125, USA }
 
\author{W.~Zheng}
\affiliation{Department of Astronomy, University of California, Berkeley, CA 94720-3411, USA}

\begin{abstract}
We present multiwavelength observations of the tidal disruption event (TDE) iPTF15af, discovered by the intermediate Palomar Transient Factory (iPTF) survey at redshift $z=0.07897$. The optical and ultraviolet (UV) light curves of the transient show a slow decay over five months, in agreement with previous optically discovered TDEs. It also has a comparable black-body peak luminosity of $L_{\rm{peak}} \approx 1.5 \times 10^{44}$\,\ergs. The inferred temperature from the optical and UV data shows a value of (3$-$5) $\times 10^4$\,K.  The transient is not detected in X-rays up to $L_X < 3 \times 10^{42}$\ergs within the first five months after discovery. The optical spectra exhibit two distinct broad emission lines in the \io{He}{ii} region, and at later times also \halpha\ emission. Additionally, emission from [\io{N}{iii}] and [\io{O}{iii}] is detected, likely produced by the Bowen fluorescence effect. UV spectra reveal broad emission and absorption lines associated with high-ionization states of \io{N}{v}, \io{C}{iv}, \io{Si}{iv}, and possibly \io{P}{v}. These features, analogous to those of broad absorption line quasars (BAL QSOs), require an absorber with column densities $N_{\rm{H}} > 10^{23}$\,cm$^{-2}$. This optically thick gas would also explain the non-detection in soft X-rays.  The profile of the absorption lines with the highest column density material at the largest velocity is opposite that of BAL QSOs.  We suggest that radiation pressure generated by the TDE flare at early times could have provided the initial acceleration mechanism for this gas. Spectral UV line monitoring of future TDEs could test this proposal.
\end{abstract}


\section{Introduction} \label{sec:intro}

Most of the galaxies in our nearby Universe contain a supermassive black hole (SMBH) at their cores \citep{KormendyRichstone1995ARAA}. Occasionally, accretion of gas in the vicinity of the SMBH reveals its presence as an active galactic nucleus (AGN). However, a large fraction of SMBHs appear to be dormant, as their activity is not directly revealed through observations in the electromagnetic spectrum. Sometimes the disruption of a star in the vicinity of the SMBH --  a tidal disruption event (TDE) -- will produce a bright flare, detectable at multiple wavelengths. This new energy release episode will shed light onto these otherwise quiescent SMBHs, allowing us to study in more detail their characteristics, such as their masses \citep{Mockler2018arXiv1} or spin distributions \citep{Stone2018arXiv}. In addition, the shocks, circularization, and accretion of the stellar debris onto the SMBH can reveal clues to better understand the more stable AGN activity or the unusual behaviour of variable quasars \citep{LaMassa2015ApJ}.

 According to theory, flares from TDEs were expected to originate in the innermost regions of the accretion disks \citep{Rees1988}, which translate into soft X-ray emission and extreme ultraviolet (UV) emission. Consequently, searches for TDEs in archival data started in X-rays using the ROentgen SATellite (ROSAT) and the XMM-Newton Slew Survey \citep{Bade1996AA,Komossa1999AA,Saxton2012AA}. Subsequently, transient UV emission was also associated with activity produced by stellar disruptions \citep{Gezari2006ApJ,Gezari2008ApJ}, and the search progressed at optical \citep{vanVelzen2011ApJ,Arcavi2014}, gamma ray, and radio wavelengths \citep{Bloom2011Sci,Burrows2011Nature,Cenko2012ApJ}. Currently, optical transient surveys are the most active discovery tool, with a total of $\sim25$ optical TDE candidates reported to date\footnote{Based on reports to the TDE Open Catalogue \url{https://tde.space}}. 
 
 The majority of recent optical TDE discoveries were monitored in the UV using broad-band photometry with the \textit{Neil Gehrels Swift Observatory} \citep[\textit{Swift};][]{Gehrels2004ApJ}. The sample has shown significantly higher temperature in the continuum than supernovae, with $T_{\rm{eff}} \approx (20-80) \times 10^3$\,K (see sample in \cite{Hung2017ApJ}), suggesting that the bulk of the energy was released in the UV, inaccessible to ground-based telescopes. To date, restframe UV spectra of only five TDEs have been published, including the object discussed herein.  While the optical spectra of these events are usually marked by \io{He}{ii}, \io{He}{i}, and occasionally H broad emission lines \citep{Holoien2016-14li,Blagorodnova2017ApJ}, far-UV spectroscopy of ASASSN14li \citep{Cenko2016ApJ}, iPTF16fnl \citep{Brown2018MNRAS}, and iPTF15af \citep[also published in][]{Yang2017ApJ} have shown the presence of broad collisionally excited lines of Ly$\alpha$, \io{C}{iv}, \io{N}{v}, and \io{Si}{iv}. In contrast, low ionization lines of \io{Fe}{ii} and \io{Mg}{ii} dominated the UV spectra for PS1-11af \citep{Chornock2014ApJ} and PS16dtm \citep{Blanchard2017ApJ}. The observed diversity suggests that UV wavelengths are especially relevant for characterizing the geometry, density, and kinematics of the stellar debris and pre-existing circumnuclear material. 
 
Here we present the discovery and follow-up observations at optical, UV, mid-infrared, X-ray, and radio wavelengths of iPTF15af, a TDE discovered by the intermediate Palomar Transient Factory \citep[iPTF;][]{Rau2009PASP} survey at redshift $z=0.07897$. We show that the photometric and spectroscopic evolution of this event resembles other tidal disruption flares reported to date. The broad absorption component observed in the UV spectrum suggests the presence of high-velocity outflows, similar to the ones observed in broad absorption line quasars (BAL QSOs). 

Section \ref{sec:obs} describes the discovery, the host galaxy, and our multiwavelength follow-up observations. Sections \ref{sec:sed_analysis} and \ref{sec:spec_analysis} describe the photometric and spectroscopic analysis of the object. In Section \ref{sec:discussion} we compare the UV spectra of iPTF15af to those of other TDEs and QSOs. In Section \ref{sec:conclusions} we summarize our conclusions.

In this work, we assume a flat cosmology and values of H$_0$ = 70 \kms\,Mpc$^{-1}$ and $\Omega_M = 0.30$, $\Omega_{\Lambda} = 0.70$.

\begin{figure}
\includegraphics[width=0.48\textwidth]{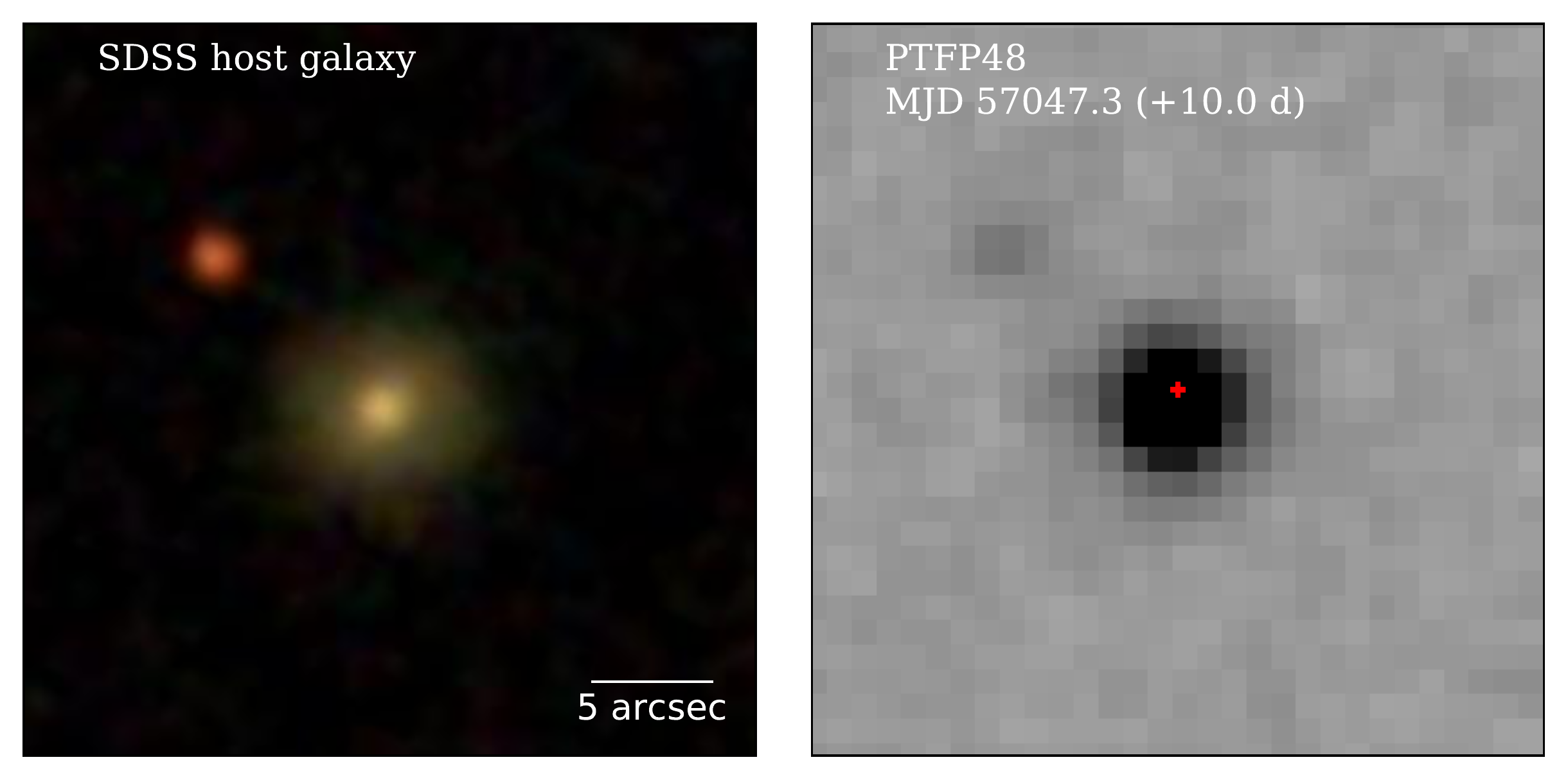} 
\caption{Left: Archival SDSS colour composite of the host galaxy. Right: iPTF P48 image with the transient at 10 days after discovery (around peak magnitude). The red cross marks the location of the transient.}
\label{fig:host}
\end{figure}

\section{Discovery and Observations}\label{sec:obs}

iPTF15af was discovered by the iPTF survey on UTC 2015 January 15.3 (MJD 57037.3). The object was detected in three of the five difference images taken on that night. The average AB magnitude \citep{OkeGunn1983ApJ} measured by the pipeline \citep{Masci2017PASP} was $R=20.7\pm0.2$ in the iPTF Mould $R$ band. Stacked forced photometry allowed us to recover the transient in archival data at 25\,days prior to its discovery in single-epoch images.

 The central location of the transient in its host galaxy (see Figure \ref{fig:host}) awarded the transient multiwavelength follow-up observations and a spectroscopic classification a week later. The spectrum confirmed the presence of strong \he2 lines, characteristic of an optical TDE candidate \citep{Gezari2012Natur,Arcavi2014}.

From the averaged positions of the residuals in the $r$ band, we derive the location of the transient to be $\alpha = 08^{\rm hr}48^{\rm m}28.12^{\rm s}$, $\delta = +22^\circ 03' 33.575''$ (J2000), with a standard deviation in both coordinates of $1\sigma=120$\,mas. This position is offset from the central location of the galaxy core by 230\,mas, which is consistent with the nucleus of the galaxy within 2$\sigma$.
The Milky Way extinction along the line of sight is $A_V=0.093$\,mag \citep{SchlaflyFinkbeiner2011ApJ}.

 \subsection{Host Galaxy} \label{sec:host}

 The transient's host galaxy SDSS J084828.13+220333.4 is located inside of the footprint of SDSS DR12 \citep{SDSSDR12}. An archival spectrum of the host was obtained on MJD 53379 as part of the SDSS spectroscopic survey. We do not identify major lines associated with star formation or AGNs. The spectroscopic redshift of the galaxy is $z=0.07897\pm 0.00004$, which corresponds to a luminosity distance $D_L=358$\,Mpc  (distance modulus $\mu=37.8$\,mag). 

Archival photometry of the host is provided in Table \ref{tab:hostmag}. The limit for the \textit{GALEX} far-ultraviolet (FUV) band was obtained from synthetic photometry of the best-fit galaxy model from the \cite{BruzualCharlot2003MNRAS} spectral library. The spectrum was reddened and normalized to the SDSS $g$-band measurement.

Literature modeling analysis of the light profile of the galaxy shows that the values derived for its bulge-to-total light ratios are $(B/T)_g=0.50$ and $(B/T)_r=0.56$ in the $g$ and $r$ bands, respectively \citep{Simard2011ApJS}. The stellar mass for the host and its bulge, log\,$(M_{*}/{\rm M}_{\odot})=10.360^{+0.099}_{-0.138}$ and  log\,$(M_{b}/{\rm M}_{\odot}) = 9.99_{-0.15}^{+0.12}$, show that $\sim 40$\% of the mass is concentrated in the bulge of the galaxy \citep{Mendel2014ApJS}. Galaxies with enhanced stellar density cores are common among TDE hosts \citep{Graur2018ApJ}, showing unusually high $B/T$ values \citep{Law-Smith2017ApJ}. High-resolution spectroscopy of the galaxy has allowed us to measure the velocity dispersion in the host, leading to an $M-\sigma$ estimate of log$_{10}\,(M_{\rm BH}/{\rm M}_{\odot}) = 6.88^{0.38}_{-0.38}$\,\citep{Wevers2017MNRAS}. 

The low specific star formation rate (sSFR) of the host, log$_{10}\,({\rm sSFR/yr})=-14.87^{+3.55}_{-0.10}$ \citep{Chang2015ApJS} shows that at the present time the galaxy is not actively forming stars. However, further analysis reveals that the host experienced a short ($\sim25$\,Myr) burst of star formation $\sim600$\,Myr ago \citep{French2017ApJ}. However, this burst could account for only a small fraction of the mass (5\%).

Since 2005, the host galaxy of the transient has been continuously monitored by the Catalina Real-time Transient Survey \cite[CRTS;][]{Drake2009ApJ}. During these last $\sim10$\,yr, the host galaxy is well detected with an average magnitude of $V=17.67\pm0.11$, which appears constant within the errors. Nuclear activity is also unlikely, as inferred from the quiescent mid-infrared (mid-IR) \textit{Wide-field Infrared Survey Explorer} \cite[\textit{WISE};][]{WISE2010} color $W1-W2=0.177$, away from the locus determined by $W1 - W2 \geq 0.8$ \citep{Stern2012ApJ}. Based on the lack of AGN features in the spectrum, the lack of variability, and no mid-IR excess, we can rule out previous AGN activity in the host at least $\sim10$\,yr before the discovery of the source.

\begin{table}
\begin{minipage}{1.\linewidth}
\begin{small}
\caption{Photometry of SDSS J084828.13+220333.4.}
\centering
\begin{tabular}{cccccccc}
\hline
Survey & Band  & Magnitude & Reference    \\ 
    &  		& (mag) &     \\ \hline
 GALEX 		& FUV$_{\rm AB}$ & $>$24.6 $^a$& [1] \\
 GALEX 		& NUV$_{\rm AB}$ & 24.02 $\pm$ 0.48 $^b$& [1] \\
 SDSSDR12 	& $u$ & 20.286 $\pm$ 0.084  $^c$& [2]\\
 SDSSDR12 	& $g$ & 18.520 $\pm$ 0.010  $^c$& [2]\\
 SDSSDR12 	& $r$ & 17.606 $\pm$ 0.007  $^c$& [2]\\
 SDSSDR12 	& $i$ & 17.186 $\pm$ 0.008  $^c$& [2]\\
 SDSSDR12 	& $z$ & 16.883 $\pm$ 0.019  $^c$& [2]\\
 
 2MASS 		& $J$ & 16.433 $\pm$ 0.132 $^d$ & [3] \\
 2MASS 		& $H$ & 15.671 $\pm$ 0.148 $^d$& [3] \\
 2MASS 		& $K$ & 15.547 $\pm$ 0.175 $^d$& [3] \\
 WISE 		& W1	& 14.809 $\pm$ 0.033 $^d$&	[4] \\
 WISE		& W2	& 14.632 $\pm$ 0.068 $^d$&	[4] 	\\
 WISE		& W3	&     $>$12.309 $^d$&	[4] 	\\
 WISE		& W4	&     $>$8.681 $^d$&		[4]  \\  
 NVSS		& 1.4\,GHz & $<$1.8 mJy& [5]\\
 ROSAT		&0.1$-$2.4 keV& $<$3.9$\times 10^{-13}\,$\ergs		& [6]\\
\hline
\end{tabular}
\begin{tablenotes}
\item[\textdagger]$^a$ Synthetic photometry. $^b$ Measured within 7.5$^{\prime\prime}$ diameter aperture. $^c$ Model magnitude. $^d$ PSF magnitude. References: [1]  \cite{Bianchi2011}, [2] \cite{SDSSDR12}, [3] \cite{2MASS_PSC_2006}, [4] \cite{WISE2010}, [5] \cite{Condon1998AJ}
 , [6] \cite{Voges1999AA}.
    \end{tablenotes}
\label{tab:hostmag}
\end{small}
\end{minipage}
\end{table}

 \subsection{Ground-Based Photometry} \label{sec:grphot}
\begin{figure*}
\includegraphics[width=\textwidth]{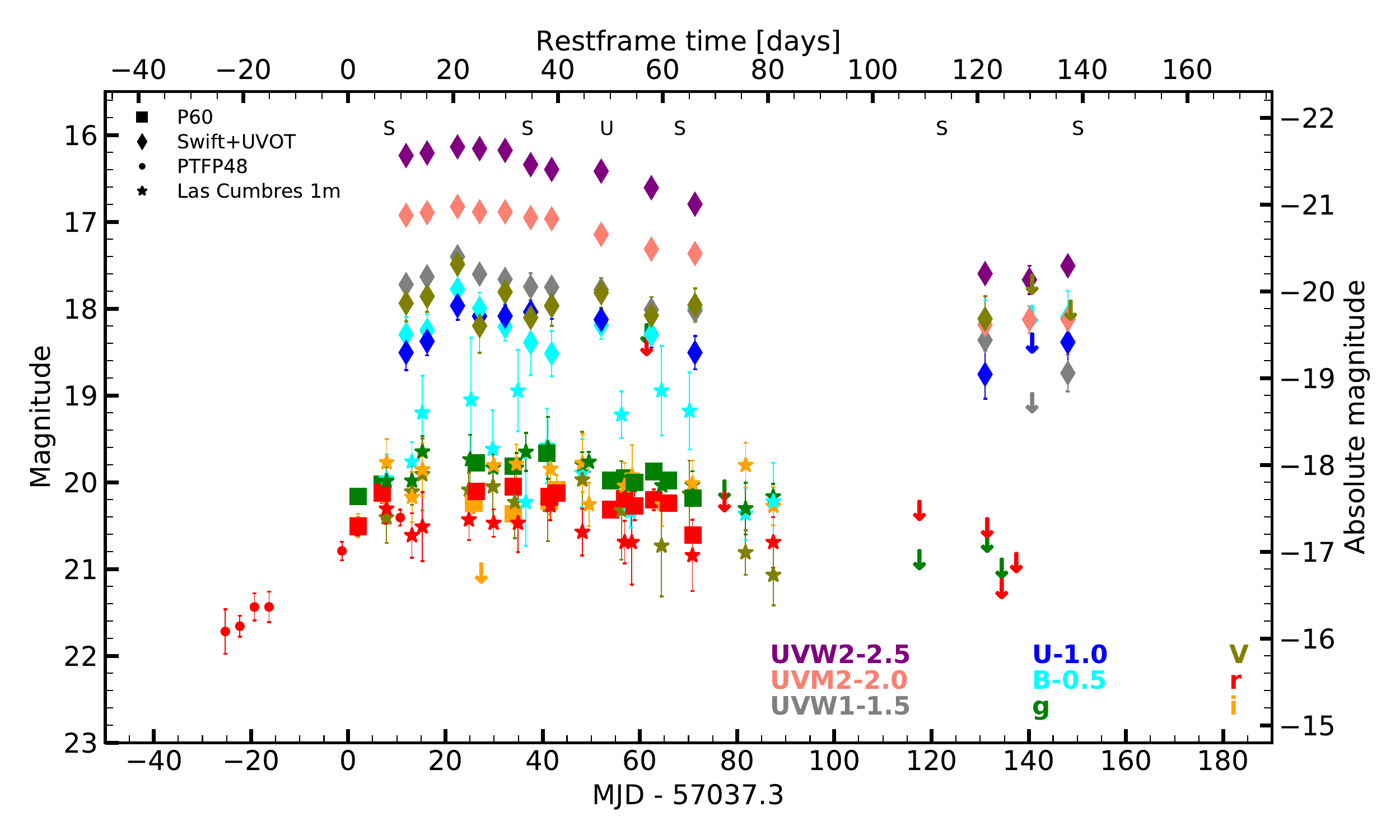} 
\caption{iPTF15af photometric measurements, corrected for Galactic extinction. For display purposes, the data have been binned with a bin size of 3\,days. PTF measurements are obtained from difference photometry. The \textit{Swift} measurements correspond to aperture photometry. Arrows show the upper limits. The epochs for optical spectra are marked with ``S'' and the epoch of the UV spectrum with ``U.'' We use the discovery date as our reference epoch. }
\label{fig:lightcurve}
\end{figure*}

 After its discovery, the transient was simultaneously monitored for 3 months in the $R$ band with the Palomar 48-inch telescope (P48) and in the $gri$ bands with the Palomar 60-inch telescope (P60). The P48 data reduction is described by \cite{Laher2014PASP}, while the photometric calibration
and the system is discussed by \citep{Ofek2012PASP}. The difference-image photometric measurements were obtained with \texttt{PTFIDE} \citep{Masci2017PASP} and \texttt{FPipe} \citep{Fremling2016} for P48 and P60, respectively. The zeropoints were calibrated using the SDSS DR13 catalogue \citep{SDSSDR13}.
 Simultaneously, we also obtained follow-up photometry in the \textit{BVgri} bands with Las Cumbres Observatory's 1\,m telescope in Texas \citep{Brown2013PASP}. The point-spread-function (PSF) photometry was obtained using \texttt{lcogtsnpipe} \citep{Valenti16}, which is calibrated to the APASS 9 \citep{Henden2016} and SDSS DR13 catalogs for \textit{BV} and \textit{gri}, respectively. Deeper, prediscovery photometric measurements were obtained by combining the flux from all available nightly sets of five images. The photometric evolution of iPTF15af is shown in Figure \ref{fig:lightcurve} and the difference-imaging measurements are available as part of Table \ref{table:phot}.

 The first noticeable characteristic of this event is its slow evolution in the UV and optical bands. Since our first detections in $r$, there is a slow brightening by $\sim1$\,mag on a timescale of 30\,days. After that, the optical bands reach a plateau phase, lasting $\sim3$ months. The observed apparent magnitudes at peak (around MJD 57370) are $m_g=19.8\pm0.1$\,mag and $m_r=20.2\pm0.1$\,mag, corresponding to absolute magnitudes (corrected for foreground extinction) of $M_g=-17.9\pm0.1$\,mag and $M_r=-17.7\pm0.1$\,mag. Because of the faint apparent magnitude of the object, our follow-up observations are non-detections in the optical bands after the fourth month; apparently, by that time the transient had faded below $\sim21$\,mag in $g$ and 22 in $r$.

\startlongtable
\begin{small}
 \renewcommand{\tabcolsep}{0.11cm}
 \begin{deluxetable*}{rrcccccccccc} 
 \tablewidth{0pt} 
 \tablecaption{Optical difference imaging and \textit{Swift} UV aperture photometry of iPTF15af in the AB magnitude system.$^a$  \label{table:phot}} 
 \tablehead{ 
   \colhead{MJD}     &  \colhead{Phase$^b$}     &\colhead{Telescope }    &  \colhead{$UVW1$} &  \colhead{$UVM2$} &  \colhead{$UVW2$} &  \colhead{$U$} &  \colhead{$B$} &  \colhead{$V$} &  \colhead{$g$} &  \colhead{$r$} &  \colhead{$i$} \\ 
\colhead{ (days) } &  \colhead{ (days) }& + Instrument & \colhead{(mag)}&  \colhead{(mag)}&  \colhead{(mag)}&  \colhead{(mag)}&  \colhead{(mag)}&  \colhead{(mag)}&  \colhead{(mag)}&  \colhead{(mag)}&  \colhead{(mag)}
} 
\startdata 
 56664.0 & $-$373.3 & PTFP48 & -- & -- & -- & -- & -- & -- & -- & $>$23.29 & --\\
 56667.0 & $-$370.3 & PTFP48 & -- & -- & -- & -- & -- & -- & -- & $>$22.20 & --\\
 56670.0 & $-$367.3 & PTFP48 & -- & -- & -- & -- & -- & -- & -- & $>$20.47 & --\\
 56769.0 & $-$268.3 & PTFP48 & -- & -- & -- & -- & -- & -- & -- & $>$21.89 & --\\
 57012.0 & $-$25.3 & PTFP48 & -- & -- & -- & -- & -- & -- & -- & 21.79$\pm$0.26 & --\\
 57015.0 & $-$22.3 & PTFP48 & -- & -- & -- & -- & -- & -- & -- & 21.73$\pm$0.12 & --\\
 57018.0 & $-$19.3 & PTFP48 & -- & -- & -- & -- & -- & -- & -- & 21.51$\pm$0.16 & --\\
 57021.0 & $-$16.3 & PTFP48 & -- & -- & -- & -- & -- & -- & -- & 21.51$\pm$0.18 & --\\
 57036.0 & $-$1.3 & PTFP48 & -- & -- & -- & -- & -- & -- & -- & 20.87$\pm$0.11 & --\\
 57039.3 & 2.0 & P60+SEDM & -- & -- & -- & -- & -- & -- & 20.27$\pm$0.06 & 20.58$\pm$0.10 & 20.55$\pm$0.14\\
 57039.0 & 1.7 & PTFP48 & -- & -- & -- & -- & -- & -- & -- & 20.64$\pm$0.05 & --\\
 57044.3 & 7.0 & P60+SEDM & -- & -- & -- & -- & -- & -- & 20.13$\pm$0.07 & 20.20$\pm$0.12 & $>$20.20\\
   ... & ... & ... & ... & ... & ... & ... & ... & ... & ... & ...\\
 \enddata 
 \vspace{0.5cm}
 $^a$The $r$-band column for the P48 contains measurements in the Mould-$R$ filter system.  These data are not corrected for Galactic extinction.  Table \ref{table:phot} is published in its entirety in machine-readable format. A portion is shown here for guidance regarding its form and content.$^b$ The phase is respect to the date of discovery MJD 57037.3.
 \end{deluxetable*}\end{small}

 \subsection{Swift Photometry} \label{sec:phot}

Following the optical discovery of iPTF15af, we initiated follow-up observations with the Ultra-Violet Optical Telescope \citep[UVOT;][]{Roming2005} on the \textit{Swift} satellite. The data were obtained in three successive campaigns in all 6 UVOT filters. We used the software package \texttt{UVOTSOURCE} to obtain the counts of the source within a $5''$ aperture radius. The background was computed from an off-target sky region using an aperture of $12''$. The magnitudes were derived using the latest UVOT calibration \citep{Poole2008,Breeveld2010MNRAS}. The aperture magnitude measurements are listed in Table \ref{table:phot}.

The evolution of the \textit{Swift} photometry is shown in Figure \ref{fig:lightcurve}. While the reddest bands, with higher contamination from host-galaxy light, follow the plateau observed in the optical bands, the UV bands $UVW2$ and $UVM2$ show a declining trend after 50\,days, which places an approximate date of the peak at $\sim40$\,days after discovery in the rest frame of the source (around MJD 57077.3). We note that, while at later times the UV emission has considerably faded, our measurements are still 4\,mag brighter than the archival ones reported by \textit{GALEX}. The flattening of UV emission at late times is consistent with the trend observed for a sample of optical TDEs, observed 5--10\,yr after the flare \citep{vanVelzen2018arXiv}.

\subsection{Optical Spectroscopy}\label{sec:opspec}

Optical spectra were obtained with the Low Resolution Imaging Spectrometer \citep[LRIS;][]{Oke1995PASP} on the Keck~I 10\,m telescope and were reduced using the automated reduction pipeline in IDL \texttt{lpipe}\footnote{\url{http://www.astro.caltech.edu/~dperley/programs/lpipe.html}}, developed by D. Perley.

The full spectral log is provided in Table \ref{tab:speclog}. Upon the publication of the article, all spectra will be made publicly available via the WISeREP repository \citep{YaronGal-Yam2012}.

\begin{figure*}
\centering
\includegraphics[width=\textwidth]{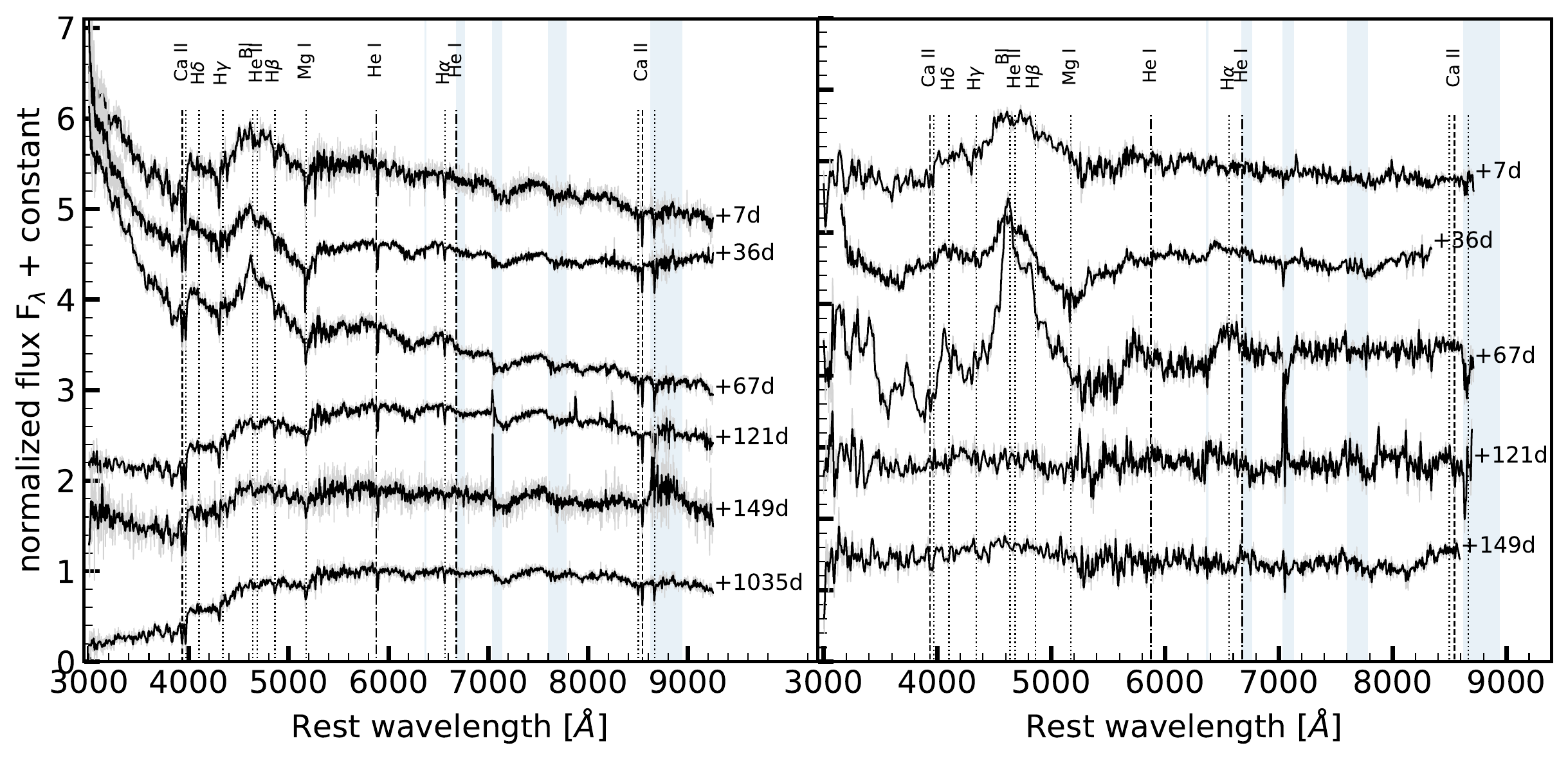}
\caption{Left: Rest-frame optical spectra of iPTF15af obtained with LRIS on the Keck~I telescope. The spectra are not corrected for Milky Way extinction. Right: Subtracted residuals for each epoch using the spectrum of the host at +1035\,days plus a black body. The broad absorption feature around 5200\,\AA\ noticeable at +36 days is related to the dichroic, which was placed at rest-frame 5600\,\AA. The most relevant element transition lines are marked.}
 \label{fig:spectra_optical} 
\end{figure*}

The spectral evolution of the event from +7\,days to +149\,days after discovery is shown in the left panel of Figure \ref{fig:spectra_optical}. In order to optimize the host-galaxy subtraction, we used the same configuration to obtain a late-time spectrum of the host at +1035\,days after discovery, when the TDE contribution was completely gone. Visual inspection of our spectrum at +149\,days after discovery confirms that for that epoch, there was still significant residual flux from the transient.

 \renewcommand{\tabcolsep}{0.21cm}
\begin{table*}
\begin{minipage}{0.9\linewidth}
\begin{small}
\caption{Log of ground-based spectroscopic observations of iPTF15af.$^a$}
\begin{center}
\begin{tabular}{rcccccccrr}
\hline
Phase$^b$ & MJD  & UTC & Telescope &     P.A.& Exposure   \\ 
  (d) &  (d) &   &	+Instrument	& 	(deg) &	(s)	     \\ \hline
+7 & 57044.5 & 2015-01-22 11:38:56 & Keck I+LRIS &  105 & 1240 \\ 
+36 & 57073.5 & 2015-02-20 12:59:35 & Keck I+LRIS & 268 & 600 \\ 
+67 & 57104.3 & 2015-03-23 07:16:37 & Keck I+LRIS &  230 & 820 \\ 
+121 & 57158.3 & 2015-05-16 06:17:20 & Keck I+LRIS &  87 & 1200 \\ 
+149 & 57186.3 & 2015-06-13 06:11:08 & Keck I+LRIS &  270 & 610 \\ 
+1035 & 58072.6 & 2017-11-15 13:33:51 & Keck I+LRIS & 272 & 1160 \\
\hline
\end{tabular}
\end{center}
$^a$All observations were conducted with a combination of grism 400/3400 and grating 400/8500, and a long 1$^{\prime\prime}$-wide slit, providing a resolution of $\sim7$\,\AA.
$^b$The phase is relative to discovery date with MJD 57037.3.
\label{tab:speclog}
\end{small}
\end{minipage}
\end{table*}

\subsection{UV Spectroscopy} \label{sec:uvspec}

\begin{figure*}
\centering
\includegraphics[width=\textwidth]{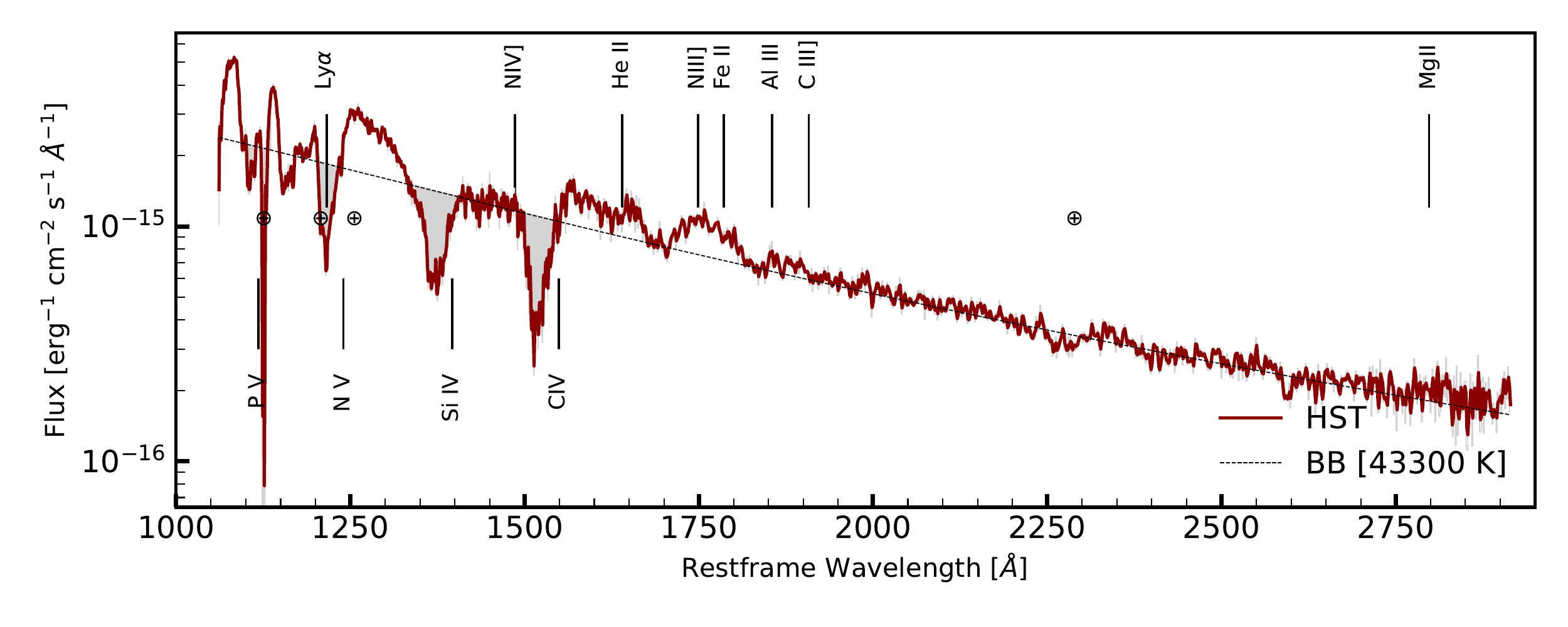} 
\caption{{\it HST}/STIS spectrum of iPTF15af at 52\,days after discovery (red dark line) and the best-fit black-body emission to the segment between 2000\,\AA\ and 3000\,\AA\ (black dashed line). The major broad emission lines have been marked in the upper part of the plot and the major broad absorption lines are shown with shaded areas in the bottom, along with a label. The spectrum has been corrected for Galactic extinction. The regions of geocoronal airglow lines are marked with $\oplus$ symbols.}
\label{fig:spectra_uv}
\end{figure*}

UV spectroscopy of iPTF15af was obtained (PI S. B. Cenko, proposal GO-13853) with the Space Telescope Imaging Spectrometer (STIS) on the \textit{Hubble Space Telescope (HST)} on UTC 2015 March 08.1 (MJD 57089.1), at 52\,days after discovery. The combined data from the far-UV (FUV) and the near-UV (NUV) MAMA detectors provide coverage in the  1145--3145\,\AA \ range. The spectrum is represented in Figure \ref{fig:spectra_uv} and the log of the UV observations is provided in Table \ref{tab:hst_log}.

 \renewcommand{\tabcolsep}{0.05cm}
\begin{table}
\begin{minipage}{1.\linewidth}
\caption{Log of UV spectroscopy of iPTF15af with HST.$^a$}
\centering
\begin{tabular}{ccccc}
\hline
UTC   & Grating & Scale & Res. power & Exposure\\ 
    &  			 & 	(\angstrom pix$^{-1}$) & ($\lambda/\Delta \lambda$) & (s)     \\ \hline
2015-03-08 01:12:13 	& G230L & 1.58  & 500$-$1010 &	2130.0 \\ 
2015-03-08 02:32:38 	& G230L & 1.58	& 500$-$1010 & 	2922.0 	\\
2015-03-08 04:08:06 	& G140L & 0.60	& 960$-$1440 & 	2922.0 	\\ 
2015-03-08 05:43:35 	& G140L & 0.60	& 960$-$1440 & 	2922.0 \\
2015-03-08 07:19:03 	& G140L & 0.60	& 960$-$1440 & 	2922.0 \\
\hline
\end{tabular}
$^a$STIS instrument and FUV-MAMA and NUV-MAMA detectors. All measurements were taken with slit 52X0.2.
\label{tab:hst_log}
\end{minipage}
\end{table}

Given the good signal-to-noise ratio (S/N) of the trace in the two-dimensional images, we used the one-dimensional spectra output by the \textit{HST} pipeline. The spectra from each detector were combined using an inverse-variance-weighted average of the one-dimensional spectra.

\subsection{X-Rays}

We used the X-Ray Telescope (XRT; \citealt{Burrows2005}) onboard the \textit{Swift} satellite to observe iPTF15af in the 0.3--10.0\,keV bandpass using the photon counting (PC) mode. The observations, with typical exposure times of $\sim2$\,ks, were taken simultaneously with the UVOT exposures. The single-epoch data only yielded non-detections, with an average 3$\sigma$ upper limit of  $(5 \pm 2) \times 10^{-3}$ counts s$^{-1}$ in this bandpass.

We measured the flux assuming that the source spectrum follows a power law with a photon index of $\Gamma=2$, and a line-of-sight  absorption in the  Milky Way of $N_{\rm{H}}=3\times 10^{20}$\,cm$^{-2}$ \citep{Willingale2013MNRAS}. 
Our count rate corresponds to an upper limit for the unabsorbed flux of $f=1.9_{-0.8}^{+0.5} \times 10^{-13}$\,erg\,cm$^{-2}$\,s$^{-1}$ in the 0.3--10.0\,keV bandpass.  At the distance of iPTF15af, this value translates to an upper limit on the X-ray luminosity of $L_{X} < 3.0\times 10^{42}$\,erg\,s$^{-1}$.

The total flux from the source in the stacked image of 33\,ks was estimated using the 99.7\% Bayesian confidence interval \citep{KraftBurrows1991ApJ}, encoded in the \texttt{astropy Python} package (assuming only a non-negativity prior). We used the exposure-weighted correction factor provided by the \textit{Swift} XRT pipeline to calculate the total of number of counts. The 3$\sigma$ level is $[0, 3.6] \times 10^{-4}$ counts s$^{-1}$, corresponding to an upper limit for the luminosity of $L_{X} \leq 2.2 \times 10^{41}$\,erg\,s$^{-1}$ in the \textit{Swift} band.

When comparing the X-ray luminosity of iPTF15af to that of other BAL QSOs, we first used the UV spectrum to derive the UV/X-ray luminosity relation, quantified by the $\alpha_{\rm{OX}}$ parameter \citep{AvniTananbaum1982ApJ,Just2007ApJ},

\begin{equation}
\alpha_{\rm{OX}} \equiv 0.3838\ \rm{log}
\left( 
\frac{\textit{f}_{\nu}(2\,{\rm{keV}}) }{\textit{f}_{\nu}(2500\,\AA)}	
\right),
\end{equation}
\noindent
where $f_{\nu}$(2\,keV) is the monochromatic unabsorbed flux density at 2\,keV, derived using the stacked image counts and a power law with photon index $\Gamma=2$, and $f_{\nu}(2500\,$\AA) is the flux density at 2500\,\AA, both in erg s$^{-1}$ cm$^{-2}$ Hz$^{-1}$.

For iPTF15af we obtained $\alpha_{\rm{OX}}=-1.8\pm0.1$ for the X-ray upper confidence level interval. In order to compare this value to that of standard QSOs, we derived the expected X-ray luminosity for a typical QSO given its UV luminosity using the relation found by \cite{Just2007ApJ}:
\begin{equation}
\alpha_{\rm{OX}}(L_{2500}) = (-0.140 \pm 0.007) \ \rm{log}(\textit{L}_{2500})+(2.705\pm 0.212).
\end{equation}


The difference between the two ratios, parametrized as
\begin{equation}
\Delta \alpha_{\rm{OX}} \equiv \alpha_{\rm{OX}} - \alpha_{\rm{OX}}(L_{2500}),
\end{equation}
\noindent
is used as an indication of X-ray weakness as compared to the general QSO population.

For our TDE, we derive a value of $\Delta \alpha_{\rm{OX}} = -0.63$, which indicates that the transient is weaker in X-rays by a factor of $\sim43$ than a typical QSO of the same UV luminosity. This value is in agreement with the weakest X-ray BAL QSOs from \cite{Gibson2009ApJ} (their Figure 16).


\subsection{Radio}

Radio follow-up observations of iPTF15af were carried out with the Jansky Very Large Array (VLA; PI A. Horesh) on UTC 2015 Jan. 31, at 15\,days after discovery.  The source was not detected either in the C (6.1\,GHz), or K (22\,GHz) bands with a root mean square (RMS) of 28 $\mu$Jy and 36$\mu$Jy respectively. The corresponding upper limits for the monochromatic radio luminosities are $\nu L_{\nu} < 2.6\times10^{37}$ \ergs and $\nu L_{\nu} < 1.2\times 10^{38}$\,\ergs at 6.1 and 22\,GHz.

These early-time observation of iPTF15af cannot completely rule out an emission mechanism equivalent to that of ASASSN-14li \citep{Holoien2016-14li}, which a month after discovery registered peak luminosities of $\nu L_{\nu}\approx9\times10^{37}$\ergs  and $\nu L_{\nu}\approx3\times10^{38}$\ergs at 5\,GHz and 15.7\,GHz, respectively \citep{vanVelzen2016Sci,Alexander2016ApJ}. In case of jetted on-axis emission, these limits would constrain the jet energy to $E_j< 10^{49}$\,erg\,s$^{-1}$ \citep{Generozov2017MNRAS}.

 \subsection{Mid-Infrared Photometry} \label{sec:mirphot}
 
We retrieved archival mid-IR data prior to and after the event from the \textit{WISE} survey and the 2017 release of \textit{NEOWISE} \citep{Mainzer2014ApJ}. The \textit{WISE} data correspond to the all-sky survey run in 3.4, 4.6, 12, and 22 $\mu$m (W1, W2, W3, W4) taken in 2010. The \textit{NEOWISE} data cover three years after the reactivation of the mission in December 2013 in the \textit{W1} and \textit{W2} bands. 

\begin{figure}[b]
\includegraphics[width=\linewidth]{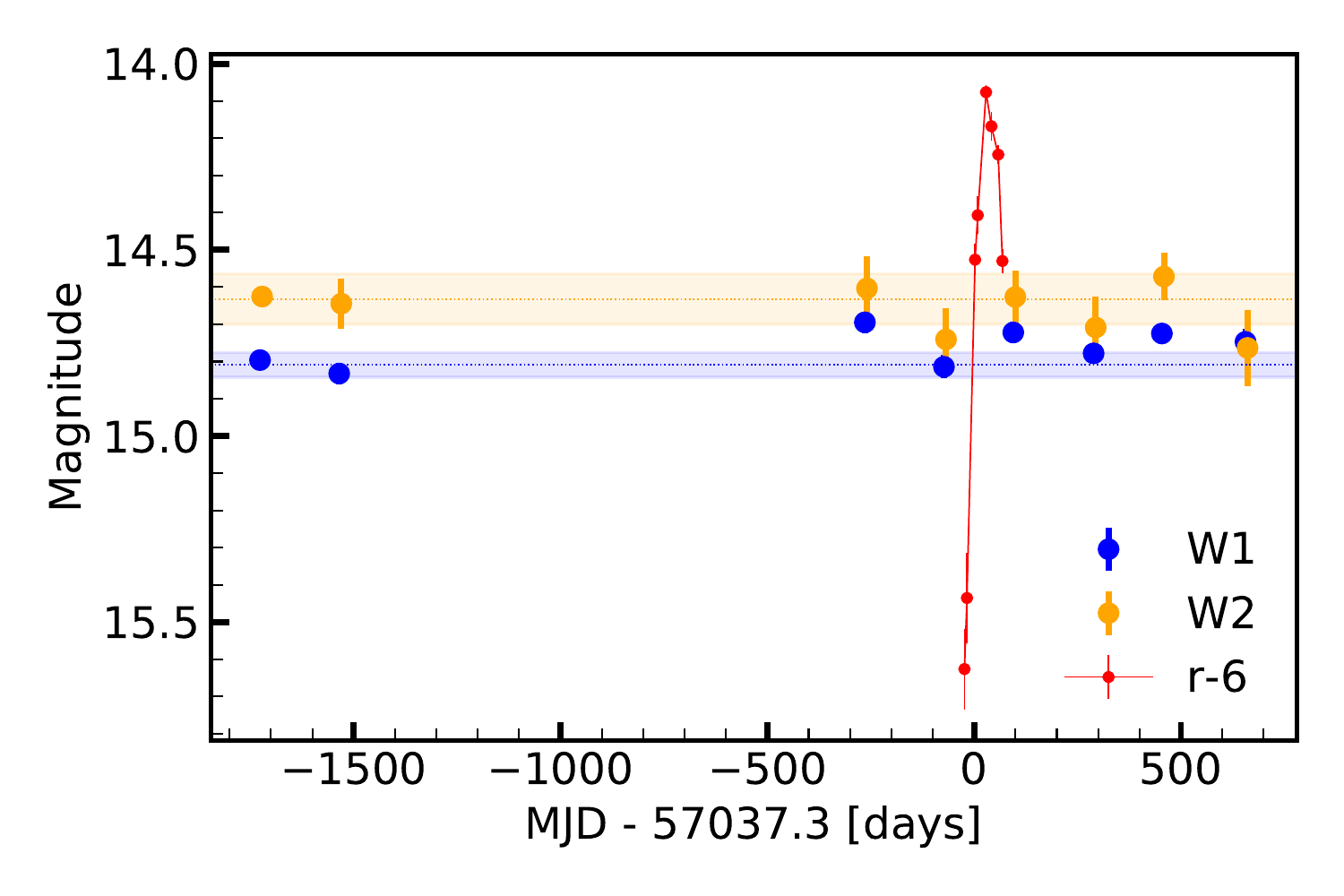} 
\caption{ \textit{WISE} (first two epochs) and \textit{NEOWISE} (last six epochs) average weighted photometry is shown with orange and blue circles. The dotted line and shaded areas show the coadded magnitude and uncertainties for the quiescent host from the \textit{AllWISE} source catalogue. The $r$-band binned photometry for the optical emission is shown with small red dots. An offset of $-$6\,mags is applied.}
\label{fig:wise_lightcurve}
\end{figure}

Using the IRSA archive\footnote{\url{http://irsa.ipac.caltech.edu}}, we selected \textit{W1} and \textit{W2} magnitudes for sources detected in single exposures within $2''$ of the position of the transient. After discarding the measurements marked as upper limits (``U'' value in the \texttt{ph\_qual} column), we obtained the flux-weighted average combining the remaining $\sim10$ measurements per epoch. The results are shown in Figure \ref{fig:wise_lightcurve}, where we also include the $r$-band light curve of the transient for comparison. 

Although the \textit{NEOWISE} light curve shows small variations around the archival value of the source in the \textit{allWISE} catalogue, none of them is more significant than 3$\sigma$. Hence, for iPTF15af we can rule out a detection of a strong IR echo as in the case of PTF09ge \citep{vanVelzen2016ApJ} or PS16dtm \citep{Jiang2017ApJ}.

 \section{Spectral Energy Distribution Analysis} \label{sec:sed_analysis}

For our spectral energy distribution (SED) analysis, we used \textit{Swift} UVOT aperture photometry for $UVW1$, $UVM2$, and $UVW2$ (subtracting the host synthetic magnitude from our best fit model spectra), and difference-image photometry in the $r$ and $g$ bands, to derive the black-body temperature and radius of the emission from the transient. We fixed the time of the \textit{Swift} observations as our reference epoch and interpolated the optical bands to derive the magnitudes at the same epoch. The fluxes, corrected for Milky Way extinction, were fit with an SED corresponding to a single-temperature black-body model. We have excluded the optical \textit{Swift} bands from our analysis, as they had considerable contamination from the host galaxy, which was still present in our last epochs. The $i$ band from the P60 exhibited larger scatter than usual, so it was excluded as well.

\begin{figure}
\includegraphics[width=0.5\textwidth]{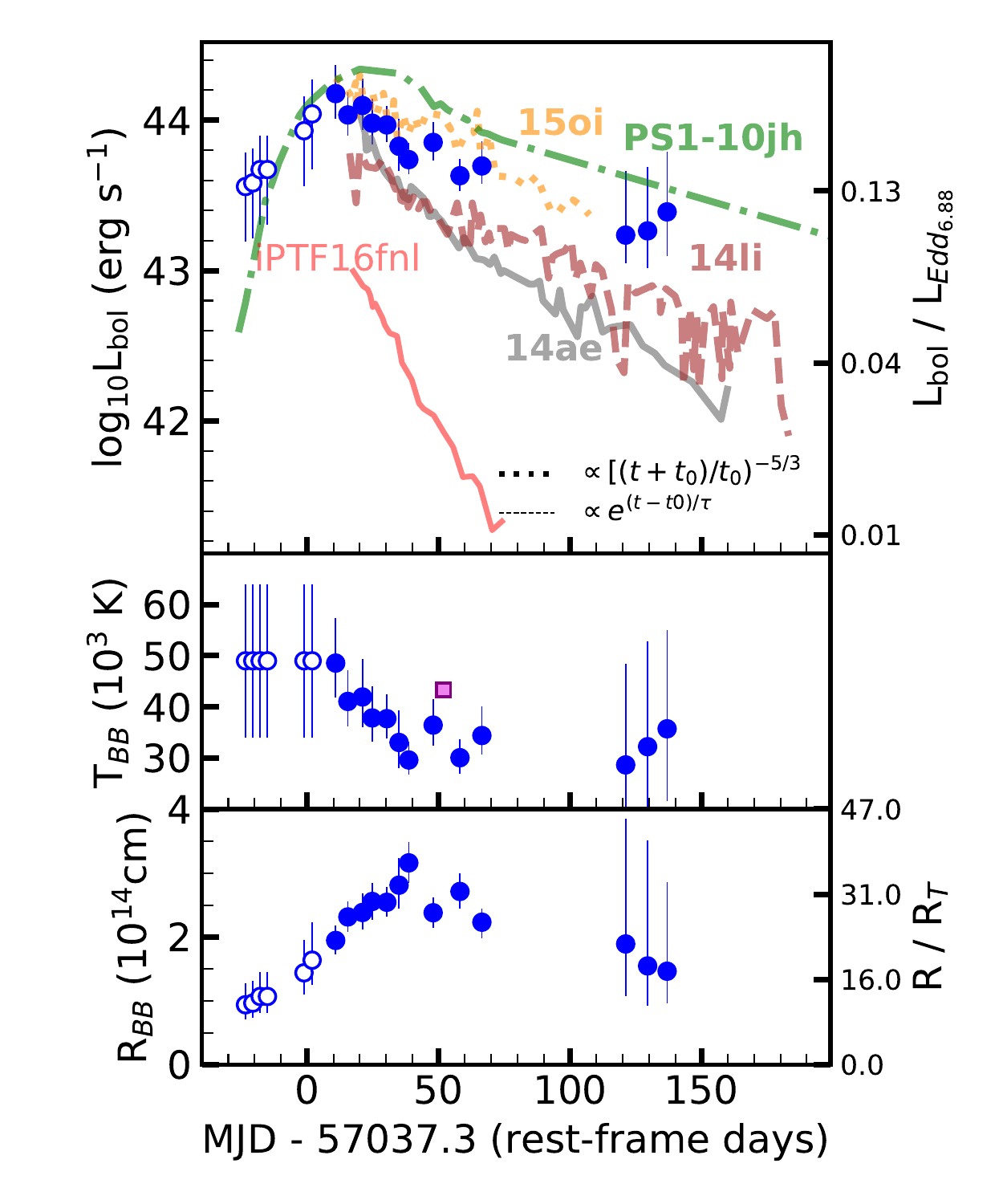} 
\caption{Top: Bolometric black-body light curve of iPTF15af. Blue filled circles represent the fits using extinction-corrected \textit{Swift} bands. Open blue circles represent the values estimated using the $r$-band scaling. The purple square shows the temperature derived using the featureless portion of the \textit{HST} UV spectrum. The right-hand top axis shows the bolometric luminosity as a fraction of the total Eddington luminosity for a $10^{6.88}$\,\Msun SMBH. Middle: Temperature evolution. Bottom: Evolution of the black-body radius. The right-hand axis shows the radius as a fraction of the tidal disruption radius for a $10^{6.88}$\,\Msun SMBH and a 1\,\Msun star. In all three plots the epoch of discovery has been adopted as the reference MJD.}
\label{fig:luminosity}
\end{figure}

The initial value for the temperature prior was derived from the fit to the UV spectrum (see Section \ref{sec:uv_spec_analysis}). Our posterior distributions for temperature and radius are based on Markov Chain Monte Carlo (MCMC) simulations, run using the \texttt{Python} package \texttt{emcee} \citep{Foreman-Mackey2013PASP}. The black-body bolometric luminosity, along with the best fit for the temperature and the radius (median, 16\% and 84\% percentiles) as derived from the SED, are shown as filled circles in Figure \ref{fig:luminosity} and given in Table \ref{tab:lumfit}. 

 \renewcommand{\tabcolsep}{0.1cm}
\begin{table}
\begin{minipage}{1.\linewidth}
\caption{Best black-body fits to the iPTF15af SED.$^a$}
\centering
\begin{tabular}{lrrrr}
\hline
MJD   & Rest epoch & $T_{\rm BB}$ & $R_{\rm BB}$ &  $L_{\rm BB}$ \\ 
(days)  & (days) & ($10^4$\,K)& ($10^{14}$\,cm) & ($10^{43}$\,erg s$^{-1}$) \\ \hline 
57012* & -23.4 & 4.90$_{-1.50}^{+1.50}$ & 0.94$_{-0.22}^{+0.34}$ & 3.62$_{-2.07}^{+2.48}$ \\
57015* & -20.7 & 4.90$_{-1.50}^{+1.50}$ & 0.97$_{-0.23}^{+0.35}$ & 3.83$_{-2.19}^{+2.62}$ \\
57018* & -17.9 & 4.90$_{-1.50}^{+1.50}$ & 1.07$_{-0.26}^{+0.39}$ & 4.70$_{-2.68}^{+3.21}$ \\
57021* & -15.1 & 4.90$_{-1.50}^{+1.50}$ & 1.07$_{-0.26}^{+0.39}$ & 4.70$_{-2.68}^{+3.22}$ \\
57036* & -1.2 & 4.90$_{-1.50}^{+1.50}$ & 1.44$_{-0.34}^{+0.52}$ & 8.51$_{-4.86}^{+5.83}$ \\
57039* & 1.9 & 4.90$_{-1.50}^{+1.50}$ & 1.64$_{-0.39}^{+0.59}$ & 11.02$_{-6.29}^{+7.54}$ \\
57049 & 10.8 &  4.86$_{-0.68}^{+0.88}$ & 1.95$_{-0.22}^{+0.23}$ & 15.03$_{-4.81}^{+8.14}$  \\
57054 & 15.5 &  4.11$_{-0.49}^{+0.62}$ & 2.32$_{-0.23}^{+0.24}$ & 10.84$_{-2.91}^{+4.58}$  \\
57060 & 21.0 &  4.19$_{-0.58}^{+0.74}$ & 2.38$_{-0.27}^{+0.30}$ & 12.54$_{-3.90}^{+6.48}$  \\
57064 & 24.7 &  3.78$_{-0.47}^{+0.62}$ & 2.56$_{-0.29}^{+0.29}$ & 9.56$_{-2.64}^{+4.34}$  \\
57070 & 30.3 &  3.77$_{-0.39}^{+0.47}$ & 2.55$_{-0.23}^{+0.24}$ & 9.30$_{-2.17}^{+3.09}$  \\
57075 & 34.9 &  3.30$_{-0.50}^{+0.62}$ & 2.81$_{-0.36}^{+0.42}$ & 6.71$_{-2.14}^{+3.49}$  \\
57079 & 38.6 &  2.96$_{-0.29}^{+0.36}$ & 3.16$_{-0.32}^{+0.33}$ & 5.47$_{-1.11}^{+1.58}$  \\
57089 & 47.9 &  3.64$_{-0.40}^{+0.50}$ & 2.38$_{-0.23}^{+0.24}$ & 7.12$_{-1.72}^{+2.65}$  \\
57100 & 58.1 &  3.00$_{-0.31}^{+0.37}$ & 2.71$_{-0.27}^{+0.29}$ & 4.27$_{-0.90}^{+1.29}$  \\
57109 & 66.5 &  3.44$_{-0.37}^{+0.56}$ & 2.23$_{-0.25}^{+0.21}$ & 4.96$_{-1.19}^{+2.30}$  \\
57168 & 121.1 &  2.86$_{-1.07}^{+1.98}$ & 1.89$_{-0.82}^{+1.96}$ & 1.72$_{-0.60}^{+2.86}$  \\
57177 & 129.5 &  3.22$_{-1.37}^{+2.06}$ & 1.55$_{-0.62}^{+1.96}$ & 1.84$_{-0.80}^{+3.07}$  \\
57185 & 136.9 &  3.57$_{-1.42}^{+1.93}$ & 1.46$_{-0.50}^{+1.39}$ & 2.46$_{-1.21}^{+3.69}$  \\

\hline
\end{tabular}
$^a$Fits to the optical and UV SED, giving the temperature ($T_{\rm BB}$), radius ($R_{\rm BB}$), and luminosity ($L_{\rm BB}$). Epochs marked with ``*" were estimated from $r$-band images only (see \S \ref{sec:sed_analysis}).
\label{tab:lumfit}
\end{minipage}
\end{table}

The rising part in our light curve only contains optical measurements, mainly in $R$-band. Our \textit{Swift} measurements start a month after the detection of the source in stacked iPTF data.  In order to  derive the luminosity and the radius of the emitting region without \textit{Swift} NUV photometry, we proceed as follows. We initially assume that the temperature of the emission does not deviate significantly from our first estimate from \textit{Swift} photometry. We use this value of $T_{\rm BB} \approx 49,000$\,K to simulate black-body emission and scale it to our $r$-band measurements. The results are shown in Figure \ref{fig:luminosity} as open circles. Their error bars indicate the change in luminosity and radius if our assumed temperature varied by $\pm15,000$\,K.

The bolometric light curve obtained for iPTF15af shows a slow rise over the initial 30 days, reaching $L_{\rm{peak}} = 1.5^{+0.8}_{-0.5} \times 10^{44}$\,\ergs and gradually decaying over the next five months, providing a rest-frame $e$-folding time of $\tau \approx 68$\,days. The overall luminosity and timescales for this TDE, as shown in Figure \ref{fig:luminosity}, agree well with those of PS1-10jh \citep{Gezari2012Natur} and slowly evolving events like ASASSN15oi \citep{Holoien2016-15oi}.

The integrated bolometric light curve over the observed interval (Figure \ref{fig:luminosity}) shows a total energy of $L \geq (8.0 \pm 1.1)\times 10^{50}$\,erg. Assuming an efficiency of $\epsilon=0.1$ for transforming the accreted mass into the observed luminosity, $E_{\rm{rad}}=\epsilon M_{acc} c^2$, the inferred lower limit for the mass of the accreted material is $M_{\rm acc}= (4.5\pm 0.6) \times 10^{-3}  $\,\Msun.

Analogous to the trend seen in other optical TDEs (see figure 11 in \cite{Hung2017ApJ}), the temperature of iPTF15af remains relatively constant for nearly 200\,days. The emission initially appears hotter, with $T_{\rm BB}\approx 49,000$\,K, and only two months later evolves toward lower temperatures in the 30,000\,K range. The cooling is related to an expansion of the photosphere from $R_{\rm BB} \approx 1 \times 10^{14}$\,cm during the rise, up to a maximum of $3 \times 10^{14}$\,cm (4300 \Rsun) during peak, placing the emission at $\sim23$ tidal radii from the SMBH. This value is similar those reported for other optical TDEs \citep{Hung2017ApJ,Wevers2017MNRAS}.

 To further constrain the characteristics of iPTF15af, we used the same \textit{Swift} \textit{UVW2}, \textit{UVM2}, and \textit{UVW1} bands, together with the difference-imaging $g$ and $r$ photometry, to model the light curve with the open-source code \texttt{MOSFiT} \citep{Guillochon2018ApJS}, based on the TDE model of \cite{Guillochon2013ApJ}. In order to avoid host-galaxy contamination, we excluded the bands $U$, $B$, and $V$.
 Provided the mass of the SMBH was available from the literature (see \S \ref{sec:host}), we used it as our prior probability, allowing our uniform distribution to vary within 3$\sigma$. 

\begin{figure}[b]
\includegraphics[width=\linewidth]{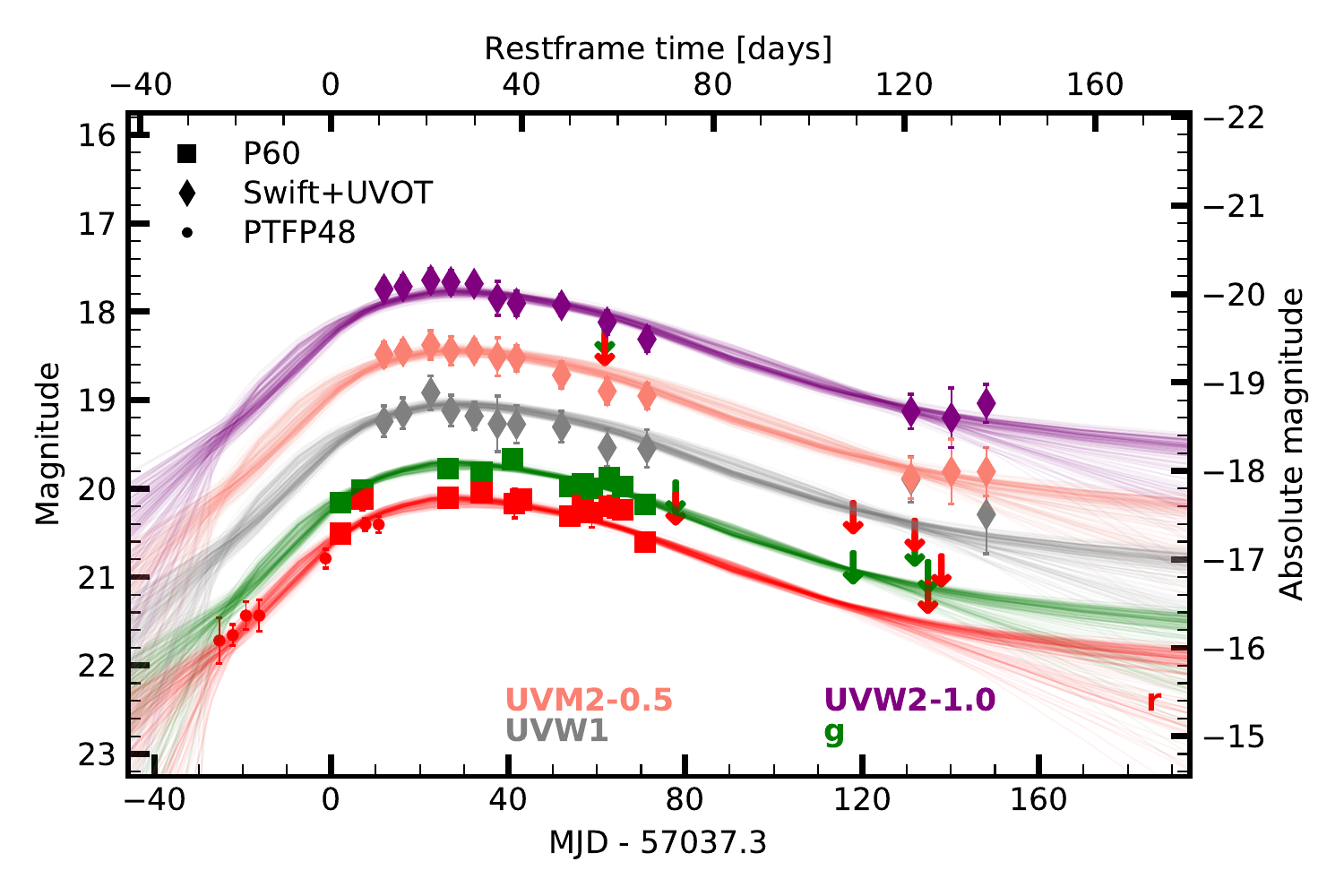} 
\caption{Markers correspond to iPTF15af optical and UV observations used for the modelling. The lines correspond to a set of best fit models, obtained with \texttt{MOSFiT}. The colour coding for the observations and the models is the same.}
\label{fig:lightcurve_model}
\end{figure}

The lightcurves for a set of best fit models are shown in Figure \ref{fig:lightcurve_model}. The parameters derived from the simulation correspond to a tidal disruption of a star with mass $M_{\rm{star}}=2.6^{+1.1}_{-0.9}$\,\Msun by a black hole with log\,$(M_{\rm BH}/\rm{M_{\odot}}) = 7.7^{+0.1}_{-0.5}$, consistent within the errors with our literature prior. This mass would be on the heavy end for the sample of TDE fits from \cite{Mockler2018arXiv1}, similar to the one computed for PS1-10jh. The estimated disruption time of the star is at 
$-38.56^{+9.6}_{-7.0}$\,days. The impact parameter, defined as the ratio between the star's disruption radius and the radius of the star at pericentre $R_T/R_p = \beta=0.10^{+0.36}_{-0.1}$, is the lowest compared to the previous sample. The value $\beta < 1$ would imply an extremely shallow encounter, corresponding to only a partial disruption \citep{Guillochon2013ApJ}. Since the event is among the brightest in the sample, its luminosity could be attributed to a highly efficient conversion of  $\dot{M}$ to luminosity of $\epsilon = 0.23^{+0.1}_{-0.1}$. The reddening in the host estimated by the model is log\,$(n_{\rm{H}}/{\rm cm}^{-2})=20.9^{+0.1}_{-0.1}$, corresponding to an extinction of $A(V) \approx 0.4$\,mag when assuming the relation provided by \cite{Guver2009MNRAS}.

 \section{Spectroscopic Analysis} \label{sec:spec_analysis}

\subsection{Optical Spectroscopy}

The optical spectroscopic evolution of iPTF15af is shown in the left panel of Figure \ref{fig:spectra_optical}. 
In the last two spectra, taken at four and five months after the event, the broad emission lines have disappeared, with the exception of a residual flux possibly associated with \halpha at +121\,days. However, the contribution from a blue continuum is still noticeable in both spectra, as compared with the quiescent host at +1035\,days. 

For our line analysis, we used the late-time (+1035\,days) spectrum of the galaxy to perform host subtraction, following the method described by \cite{Blagorodnova2017ApJ}. To ensure that the TDE features were completely gone, we compared our spectrum to the archival SDSS pre-flare spectrum, taken $\sim$13\, years before the detection of the transient. Except a small scaling factor, we found no significant deviation between the two. We then fit simultaneously the host and a black-body component to the TDE emission spectrum and subtracted this fit from the transient spectrum. A low-order polynomial was used to level up any leftover broad component in the continuum. The residuals are shown in the right-hand panel of Figure \ref{fig:spectra_optical}. 

\begin{figure}
\includegraphics[width=0.5\textwidth]{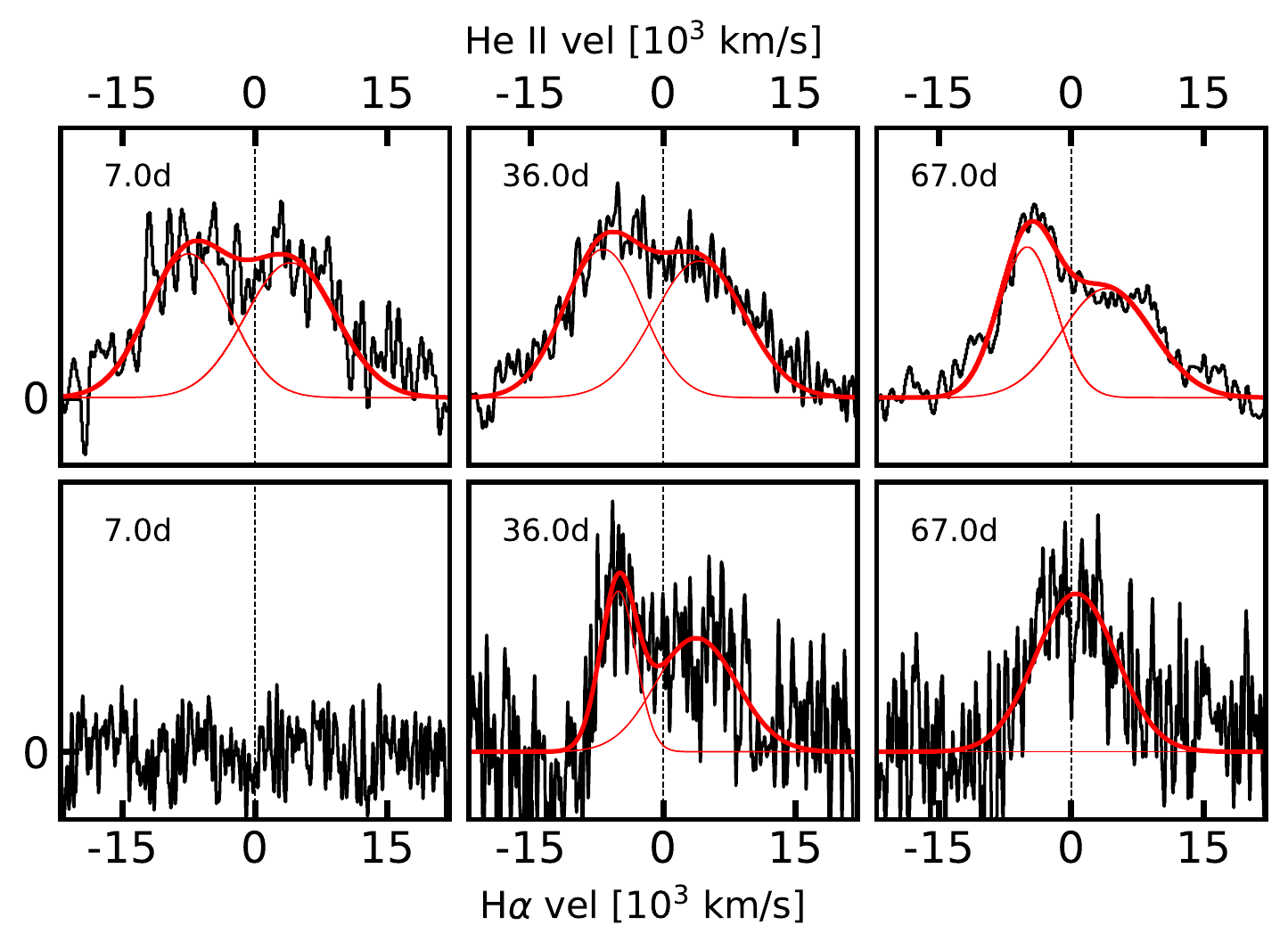} 
\caption{Best-fit line profiles to the residual emission component in iPTF15af for \io{He}{ii} $\lambda$4686 (top) and \halpha (bottom). The center of the emission lines for each element has been placed at zero velocity. The flux density has been normalized for visualization purposes.}
\label{fig:optical_fit}
\end{figure}

Using the \texttt{Python} package \texttt{lmfit}, we model the emission lines of the residual spectra with one and two Gaussian profiles and a broad exponential profile for the local continuum. The best fits are shown in Figure \ref{fig:optical_fit} and the measurements are logged in Table \ref{tab:bal_optical}.

\begin{figure*}
\centering
\includegraphics[width=\textwidth]{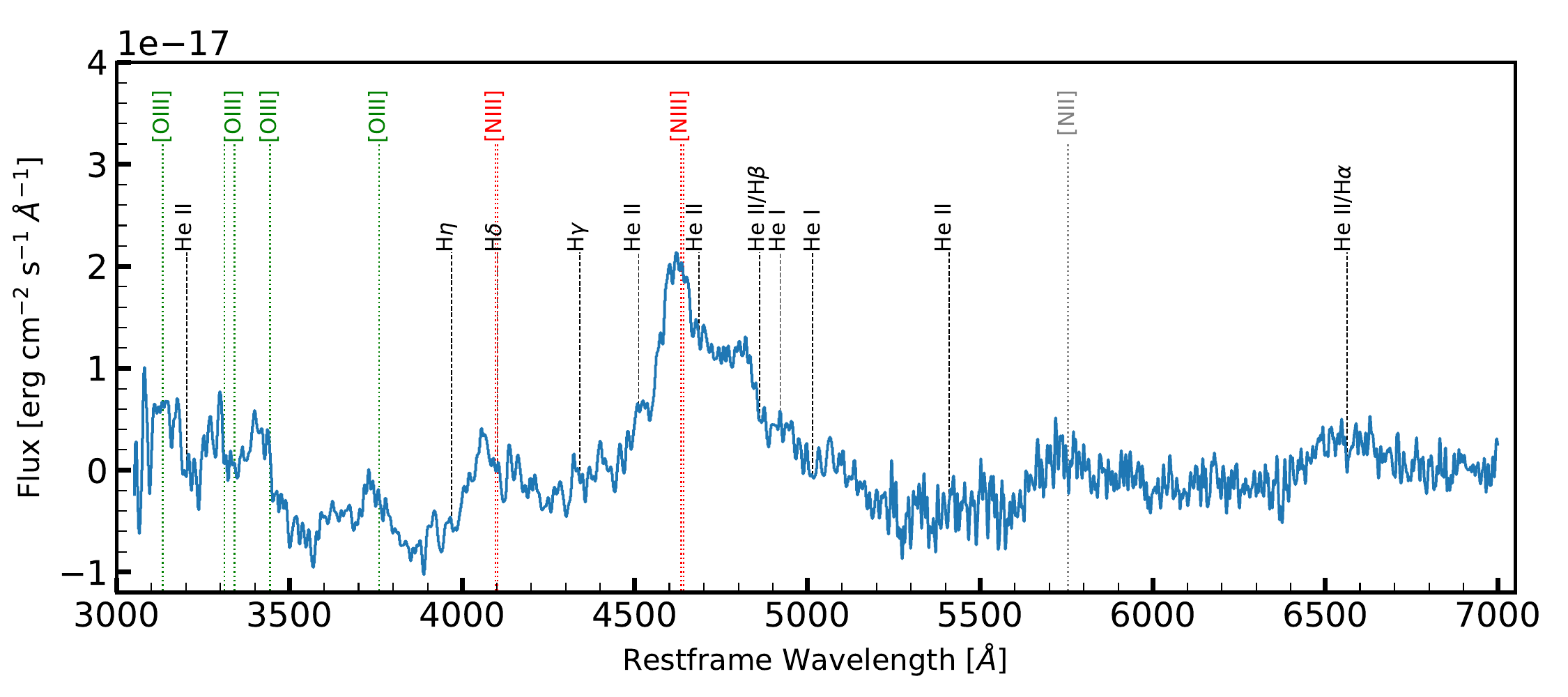} 
\caption{Identification of Bowen fluorescence lines in the spectrum at +67\,days. The [\io{N}{iii}] lines are marked in red and [\io{O}{iii}] lines in green. The centroids of these forbidden lines appear shifted by $-3000$\,\kms.}
 \label{fig:bowen_lines} 
\end{figure*}

The \halpha line is not detected in our first residual spectrum. However, at later epochs emission is present with possibly two distinct components at +36\,days, with velocities similar to those of the \he2 line. Toward +67\,days, it evolves into a single broad ($\sim 11,000$\,\kms) line centered on its rest-frame wavelength.

The \io{He}{ii} $\lambda$4686 line is clearly detected in all three spectra, which show a bimodal profile. While the redshifted wing has constant $v\approx4000$\,\kms and a full width at half-maximum (FWHM) of $\sim12,000$\,\kms, there is a noticeable narrowing of the blueshifted component at later times. The velocity shifts toward lower values and the emission becomes narrower, evolving from an average FWHM of $\sim 11,000$\,\kms to 7500\,\kms.  

The blueshifted component in the double-peaked \he2 region is very interesting. Previous spectroscopic studies of optical TDEs attributed the blueshifted component to a blend of \io{C}{iii}/\io{N}{iii}, usually detected in Wolf Rayet stars and supernovae having high temperatures \citep{Gezari2015ApJ,Brown2018MNRAS}. Figure \ref{fig:bowen_lines} shows that the centroid of this line appears to be blueshifted with respect to the rest-frame wavelengths of \io{N}{iii} $\lambda4641$ and \io{C}{iii} $\lambda4649$. Here we propose that this line and other emission features blueward of 4500\,\AA\ are caused by the Bowen fluorescence mechanism.

The Bowen fluorescence mechanism \citep{Bowen1934PASP,Bowen1935ApJ} was initially proposed to explain the unusually strong metallic lines observed in planetary nebulae. The source of the emission lies in the recombination of fully ionized \io{He}{ii} and its de-excitation toward its ground state. While the jumps between the outer orbits produce optical emission (such as the \io{He}{ii} $\lambda$4686 line), the final transition corresponding to the \he2 Ly$\alpha$ line produces an extreme-UV 303.780\,\AA\ photon. After scattering within the nebula, this photon can be either reabsorbed by another \io{He}{ii} atom, or enter in resonance with the  \io{O}{iii} lines at $\lambda$303.800 (O1) and $\lambda$303.695 (O3). The de-excitation of these lines produces the primary and  secondary Bowen decays in the optical, following a tertiary decay emitting a 374.436\,\AA\ photon.

The \io{O}{iii} line photons created here can easily escape from the nebula, resulting in transitions at $\lambda\lambda$3047, 3133, 3312, 3341, 3444 and 3760. However, the extreme UV photons are likely to be reabsorbed, this time by the resonance \io{N}{iii} line, with transitions at 374.434 and 374.442\,\AA. The recombination of the nitrogen atom toward its ground state is converted into a primary transition at $\lambda$4640 and a secondary one at $\lambda$4100, along with a 452\,\AA\ photon. While the extreme-UV photons are optically thick and suffer multiple absorptions and re-emissions within the nebula, the optical photons will easily escape.  Observations of planetary nebulae suggest that high optical depths are required for the Bowen fluorescence to work \citep{Selvelli2007AA}, as the \io{O}{iii} photons would need to suffer several scatterings in order to increase the pumping efficiency for \io{N}{iii}.

Bowen fluorescence was predicted to occur also around accreting black holes several decades ago \citep{Netzer1985ApJ}. However, only recently it has been robustly identified in TDEs (present work) and AGN flares \citep{Trakhtenbrot2019NatAs}.

Figure \ref{fig:bowen_lines} shows the main Bowen transition lines overplotted on the spectrum taken at +67\,days. In the earlier spectra at +7\,days and +36\,days we can also identify the lines associated to nitrogen $\lambda\lambda$4100, 4640. However, the broader line profile at those epochs makes the identification of the oxygen lines challenging.

In addition, in the early-time spectra we tentatively identify an emerging line around 5700\,\AA, close to the \io{He}{i} $\lambda$5875 region. However, the origin of this line is less clear, as its blueshift velocity would be around $-9000$\,\kms, higher than the one observed for \he2 and \halpha ($-7500$ to $-5000$\,\kms). Provided other nitrogen lines were detected, another possible candidate would be the \io{N}{ii} forbidden line, with a wavelength of 5754.8\,\AA.

\hspace{-0.5cm}
\renewcommand{\tabcolsep}{0.05cm}
\begin{small}
\begin{table}
\caption{Fit parameters for the optical broad lines.$^a$ \label{tab:bal_optical}}
\begin{tabular}{rlcccc} 
\hline
Phase & Ion  & Velocity$_{1}$ &FWHM$_{1}$ & Velocity$_{2}$ & FWHM$_{2}$ \\
(day)  & 	& ($10^3$\,\kms)	& ($10^3$\,\kms)	& ($10^3$\,\kms)& ($10^3$\,\kms) \\ \hline
+7  & \io{He}{ii} 		& $-$7.4$\pm$2.1 & 11.0$\pm$1.2 & 4.1$\pm$2.6 & 12.1$\pm$0.7 \\
+36  & \halpha 		& $-$5.1$\pm$0.1 & 4.7$\pm$0.3 &  3.8$\pm$0.3 & 10.8$\pm$1.1\\
+36  & \io{He}{ii} 	& $-$6.8$\pm$0.3 & 10.4$\pm$0.4 & 4.1$\pm$0.1 & 12.1$\pm$0.1  \\
+67 & \halpha 		& 0.5$\pm$0.2 & 10.8$\pm$0.5 & ... & ... \\
+67 & \io{He}{ii} 	& $-$4.9$\pm$0.2 & 7.6$\pm$0.3 & 4.1$\pm$0.4 & 12.1$\pm$0.3 \\ \hline
\end{tabular}
$^a$The index ``1'' indicates the blueshifted component and the index ``2'' the redshifted one.
\end{table}
\end{small}

\subsection{UV Spectroscopy} \label{sec:uv_spec_analysis}

The UV spectrum of iPTF15af, corrected for Milky Way extinction, is shown in Figure \ref{fig:spectra_uv}. Assuming negligible UV flux contribution from the host galaxy, we use the featureless part of the spectrum with $\lambda \geq 2000$\,\AA\ to fit a black-body spectrum. Our best fit provides a temperature of $43,300^{+1,700}_{-1,500}$\,K, consistent with the value derived from our \textit{Swift} SED analysis at a similar epoch, as shown in Figure \ref{fig:luminosity}.

The spectrum exhibits a combination of narrow absorption lines of low-ionization elements both in the Milky Way and in the TDE host galaxy, and broad lines associated with highly ionized outflows in the host. The description of each set of features is detailed below.

\subsubsection{Milky Way Absorption}

\begin{figure*}
\includegraphics[width=\textwidth]{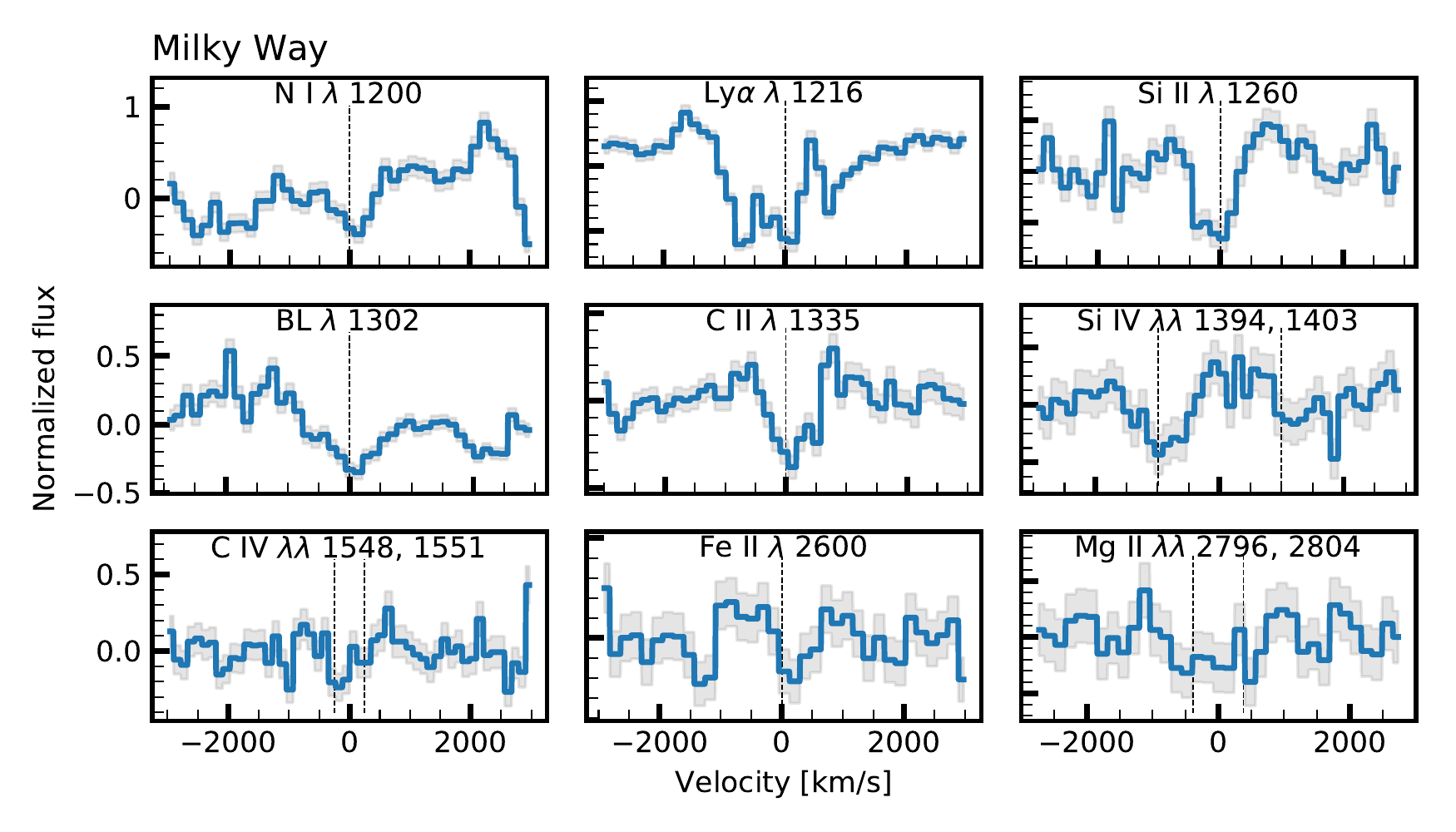} \\
\includegraphics[width=\textwidth]{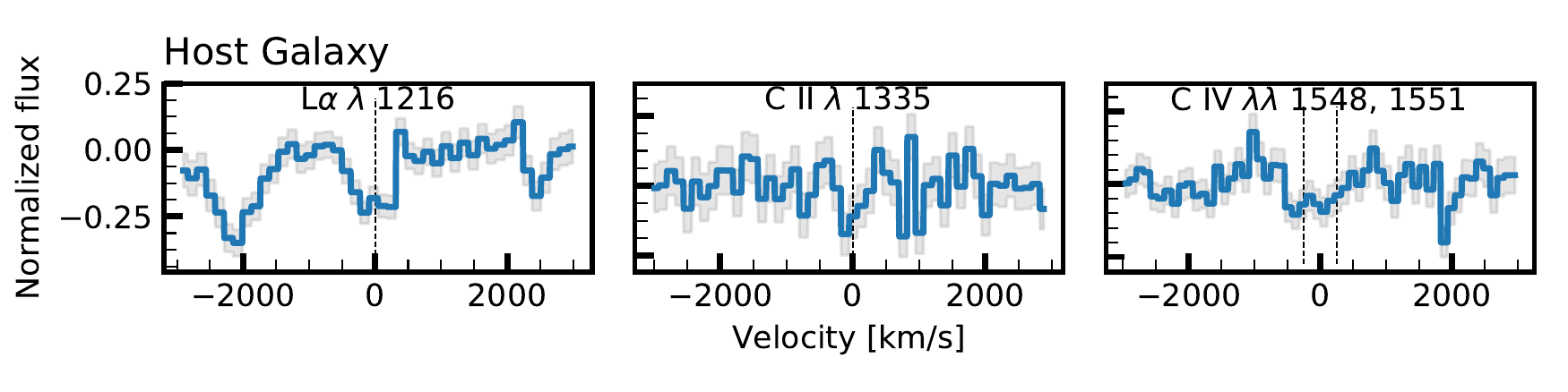} 
\caption{Top: Lines identified as part of the Milky Way ISM. The velocity was assumed to be zero at the rest-frame wavelength of each line. Vertical dashed lines show the velocity of each line. The grey shaded area shows the spectrum 1$\sigma$ uncertainty. Bottom: Lines identified at the redshift of the host galaxy. 
}
\label{fig:uv_narrow}
\end{figure*}

Figure \ref{fig:uv_narrow} shows the line identification for elements close to their rest-frame values. Although the uncertainties in the flux appear to be small, several lines show signs of blending, biasing the line-center measurements. The average resolution of our FUV ($\sim300$\,\kms) and NUV ($\sim400$\,\kms) spectra makes the precise identification of nearby absorption lines challenging. Tentatively, we identified several transitions associated with the Galactic interstellar medium (ISM), containing low-ionization elements \io{N}{i} (14.5\,eV), Ly$\alpha$ (13.6\,eV), \io{Si}{ii} (16.3\,eV), \io{C}{ii} (24.4\,eV), \io{Fe}{ii} (16.2\,eV), and possibly \io{Mg}{ii} (15.0\,eV). We identify and mark in the spectrum in Figure \ref{fig:spectra_uv} the geocoronal airglow lines of Ly$\alpha$, \io{O}{i} $\lambda$1304, \io{O}{i]} $\lambda$1356, and [\io{O}{ii}] $\lambda$2471. The high-ionization metal states correspond to \io{Si}{iv} (45.1\,eV) and \io{C}{iv} (64.5\,eV). Around 1302\,\AA\ we identify an absorption corresponding to a blend of \io{O}{i} $\lambda$1302 and \io{Si}{ii} $\lambda$1304 lines (identified as BL).

 \renewcommand{\tabcolsep}{0.05cm}
 
\begin{table}
\centering
\caption{Narrow absorption in UV spectra of iPTF15af. \label{tab:narrow_absorption}}
\begin{tabular}{lcccccr}
\hline
Ion & $\lambda_0$ & z & $\lambda_{\rm obs}$ & FWHM & EW & velocity \\
 & (\AA) &  & (\AA) & (\AA) & (\AA) & (\kms)\\ \hline
N I & 1200.0 & 0.0 & 1200.3$\pm$0.2 & 2.4 & 0.32 & 75$\pm$50\\
Ly$\alpha$ & 1215.9 & 0.0 & 1214.68$\pm$0.3 & ... & ... & -300$\pm$74\\
Si II & 1260.4 & 0.0 & 1260.06$\pm$0.1 & 2.6 & 0.7 & $-$81$\pm$24\\
BL & 1302.0 & 0.0 & 1303.0$\pm$0.3 & 2.4 & 0.2 & 230$\pm$70\\
C II & 1334.5 & 0.0 & 1335.0$\pm$0.1 & 3.21 & 0.90 & 112$\pm$22\\
Si IV & 1393.8 & 0.0 & 1394.2$\pm$0.1 & 3.86 & 0.54 & 86$\pm$22\\
C IV & 1548.2 & 0.0 & 1548.5$\pm$0.3 & 1.13 & 0.38 & 58$\pm$58 \\
Fe II & 2600.2 & 0.0 & 2602.0$\pm$0.3 & 4.7 & 1.20 & 208$\pm$35\\
Mg II & 2803.5 & 0.0 & 2801.1$\pm$0.3 & ... & ... & $-$257$\pm$32\\ \hline

Ly$\alpha$ & 1215.7 & 0.07897 & 1311.6$\pm$0.2 & 2.65 & 0.96 & $-$100$\pm$50 \\
C II & 1334.5 & 0.07897 & 1439.4$\pm$0.4 & ... & ... & $-$90$\pm$90\\
C IV & 1548.2 & 0.07897 & 1669.7$\pm$0.6 & ... & ... & $-$140$\pm$110\\
C IV & 1550.8 & 0.07897 & 1672.2$\pm$0.7 & ... & ... & $-$200$\pm$ 125\\ \hline
\end{tabular}
\end{table}

\subsubsection{Host-Galaxy Absorption}

 In agreement with the UV spectroscopic signature for ASASSN-14li, the spectrum of iPTF15af also shows weak Ly$\alpha$ absorption from the host. The majority of low-ionization elements are also missing. 
Some narrow absorption lines that were identified correspond to Ly$\alpha$ at 1216\,\AA, \io{C}{ii} $\lambda$1334,  and higher-ionization line corresponding to the \io{C}{iv} $\lambda\lambda$1548, 1551 doublet (64.5\,eV). The \io{N}{v} $\lambda\lambda$1239, 1243 doublet, if present, is likely blended into the \io{C}{ii} line at 1335\,\AA\ at Galactic redshift. We also measure a slight blueshift in our lines,  with velocities of 90--200\,\kms. These values would be consistent with the motion of gas inside of the host galaxy. However, there is a chance of a low-velocity outflow in this TDE as well, similar to the one observed for ASASSN-14li \citep{Miller2015Natur}.

 \subsubsection{Broad Lines}

According to the classical definition by \cite{Weymann1981ARAA}, a BAL is a trough at $\geq10$\% below the continuum level which extends for more than 2000\,\kms. This definition has been formalized by the so-called ``balnicity index'' \citep{Weymann1991ApJ}, computing the equivalent width (EW) of the broad ($\geq 2000$\,\kms) absorption-line troughs for each element. Using the best-fit black-body spectrum as our continuum level, we compute that iPTF15af has balnicity indices of 3856 for \io{Si}{v}, 3495 for \io{N}{v}, and 3392 for \io{C}{iv}, consistent with the distribution observed for BAL QSOs \citep{Gibson2009ApJ}. The main BALs in the spectrum of iPTF15af have been identified and marked in Figure \ref{fig:spectra_uv}.

We also identified a feature in the region around 1100\,\AA, known to contain the resonance multiplet of \io{Fe}{iii} (UV1), with a rest wavelength of 1122.5\,\AA. However, based on the similarity between the spectra of iPTF15af and high-ionization BAL QSOs (HiBAL), we attribute the broad depression in that region to \io{P}{v} $\lambda\lambda$1118, 1128 (65.0\,eV), observed in 3.0--6.2\% of BAL QSOs \citep{Hamann1998ApJ,Capellupo2017MNRAS}.  


Figure \ref{fig:spectra_uv} shows several broad emission lines (BEL) that were also identified at the host redshift. We detect strong emission from \io{N}{iii]} $\lambda$1759 (47.5\,eV) and also \io{He}{ii} $\lambda$1641 (54.4\,eV). The resonance line transitions produced by \io{C}{iv} $\lambda \lambda$1548, 1551 (64.5\,eV), \io{Si}{iv} $\lambda \lambda$1394, 1403 (45.1\,eV), and \io{N}{v} $\lambda \lambda$1239, 1243 (97.9\,eV) are also present, accompanied by broad blueshifted absorption analogous to BAL QSOs.  

Similar to other TDEs in quiescent galaxies \citep{Cenko2016ApJ,Brown2018MNRAS}, we notice the lack of emission for the low-ionization lines of \io{Mg}{ii} $\lambda\lambda$2796, 2804, \io{Al}{iii} $\lambda\lambda$1854, 1862, and \io{Fe}{ii}. Furthermore, \io{C}{iii]} $\lambda$1908, having a similar ionization potential to that of \io{N}{iii]} (47.9\,eV vs. 47.5\,eV), is also absent. Since  the ratio \io{C}{iii]}/\io{N}{iii]} has only a moderate dependence on the physical conditions of the gas (under local thermodynamic equilibrium conditions), the lack of C in the spectra of TDEs has been attributed to the unusually high abundance ratio of N to C in the debris of the disrupted star \citep{Kochanek2016MNRAS,Yang2017ApJ}.

Table \ref{tab:bal} shows the bulk velocity and the FWHM of each broad line, as modeled with a combined absorption and emission Gaussian line profiles. \io{He}{ii} and \io{N}{iii}] are only detected in emission, with their centroids redshifted by $\sim 2600$\,\kms and 100\,\kms, respectively. The observed widths of the emission lines are in the $ 10^4$\,\kms regime, corresponding to a virialized gas orbiting a log\,$(M_{\rm{BH}}/ {\rm M}_{\odot})=6.88$ SMBH at a distance of $\sim 1 \times 10^{15}$\,cm. This value is four times larger than our inferred black-body photospheric radius, suggesting that the line formation region is located likely in an outflow outside the continuum emitting zone. 

\begin{figure}
\hspace{-0.1cm}
\includegraphics[width=0.5\textwidth]{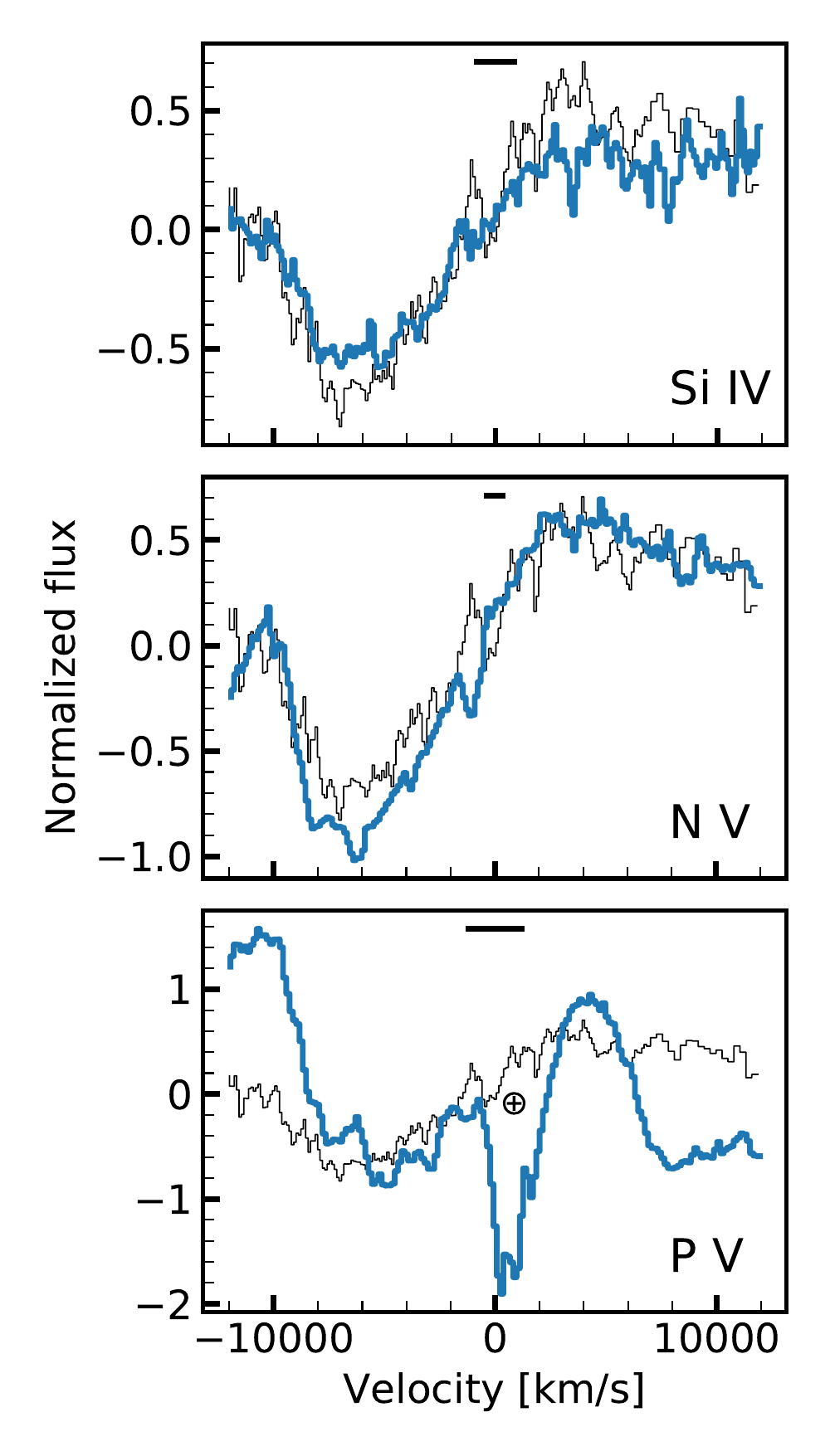} 
\caption{Comparison of the BAL of \io{Si}{iv} $\lambda \lambda$1394,1403, \io{N}{v}, $\lambda \lambda$1239,1243, and \io{P}{v} $\lambda \lambda$1118,1128 lines with the \io{C}{iv} $\lambda\lambda$1548, 1551 lines. The black line on the top indicates the separation in velocity of the doublet components. In the \io{P}{v} panel, the symbol $\oplus$ indicates absorption lines associated with the Milky Way.}
\label{fig:uv_bal}
\end{figure}

In Figure \ref{fig:uv_bal}, we compare the normalized absorption profile of \io{C}{iv} to the other high-ionization broad lines. The velocity shifts for broad absorption components in iPTF15af are $\sim -5000$\,\kms, consistent with the outflow velocities of 5000--10,000\,\kms observed in BAL QSOs. Despite some differences in the level of continuum, the widths of the absorption troughs are consistent with each other, suggesting that the absorption is produced by the same cloud. 

Assuming that the outflow in iPTF15af was initiated around the bolometric peak light (as observed in the spectral signature of PS1-11af), we can estimate the distance traveled by the cloud until the moment the UV spectrum was taken ($\sim$30\,days). Considering the outflow's fastest absorption component traveling at $\sim$8000\,\kms, we can deduce an upper limit for the distance of the absorber of $\sim 2 \times 10^{15}$\,cm, which is in agreement with the virial radius computed from the emission lines width.

\renewcommand{\tabcolsep}{0.1cm}
\begin{table}
\begin{small}
\centering
\caption{Fit parameters for the UV broad lines.$^a$}
\label{tab:bal}
\begin{tabular}{lcccc}
\hline
Ion & Velocity$_{\rm abs}$ & FWHM$_{\rm abs}$ & Velocity$_{\rm em}$ & FWHM$_{\rm em}$ \\ 
 & ($10^3$\,\kms) & ($10^3$\,\kms) & ($10^3$\,\kms) & ($10^3$\,\kms) \\ \hline
\io{Si}{iv} & $-$6.1$\pm$0.5 & 11.0$\pm$2.2 & ... & ...\\
\io{N}{iii]}& ... & ... & $-$0.2$\pm$0.3 & 10.8$\pm$1.0 \\
\io{He}{ii} & ... 				& ... 		& 2.6$\pm$0.4 & 4.1$\pm$0.9 \\
\io{C}{iv} 	& $-$5.3$\pm$0.5 & 9.1$\pm$1.4 & $-$1.6$\pm$6.2 & 14.0$\pm$4.0\\
\io{P}{v} 	& $-$4.60$\pm$0.2 & 7.4$\pm$0.8 & ... & ...\\
\io{N}{v} 	& $-$5.1$\pm$0.2 & 7.7$\pm$0.5 & 4.2$\pm$0.1 & 6.1$\pm$0.5 \\  \hline

\end{tabular}
$^a$Identified in iPTF15af. The suffix {\rm em} symbolizes the values corresponding to the emission and {\rm abs} to the absorption components. The elements are ordered from lower to higher ionization potentials: 
\io{Si}{iv} $-$ 45.1\,eV, 
\io{N}{iii]} $-$ 47.5\,eV,
\io{He}{ii}  $-$ 54.4\,eV, 
\io{C}{iv}  $-$ 64.5\,eV,
\io{P}{v}  $-$ 65.0\,eV, and 
\io{N}{v}  $-$ 97.9\,eV.
\end{small}
\end{table}

\section{Discussion} \label{sec:discussion}

\begin{figure*}[t]
\centering
\includegraphics[width=0.9\linewidth]{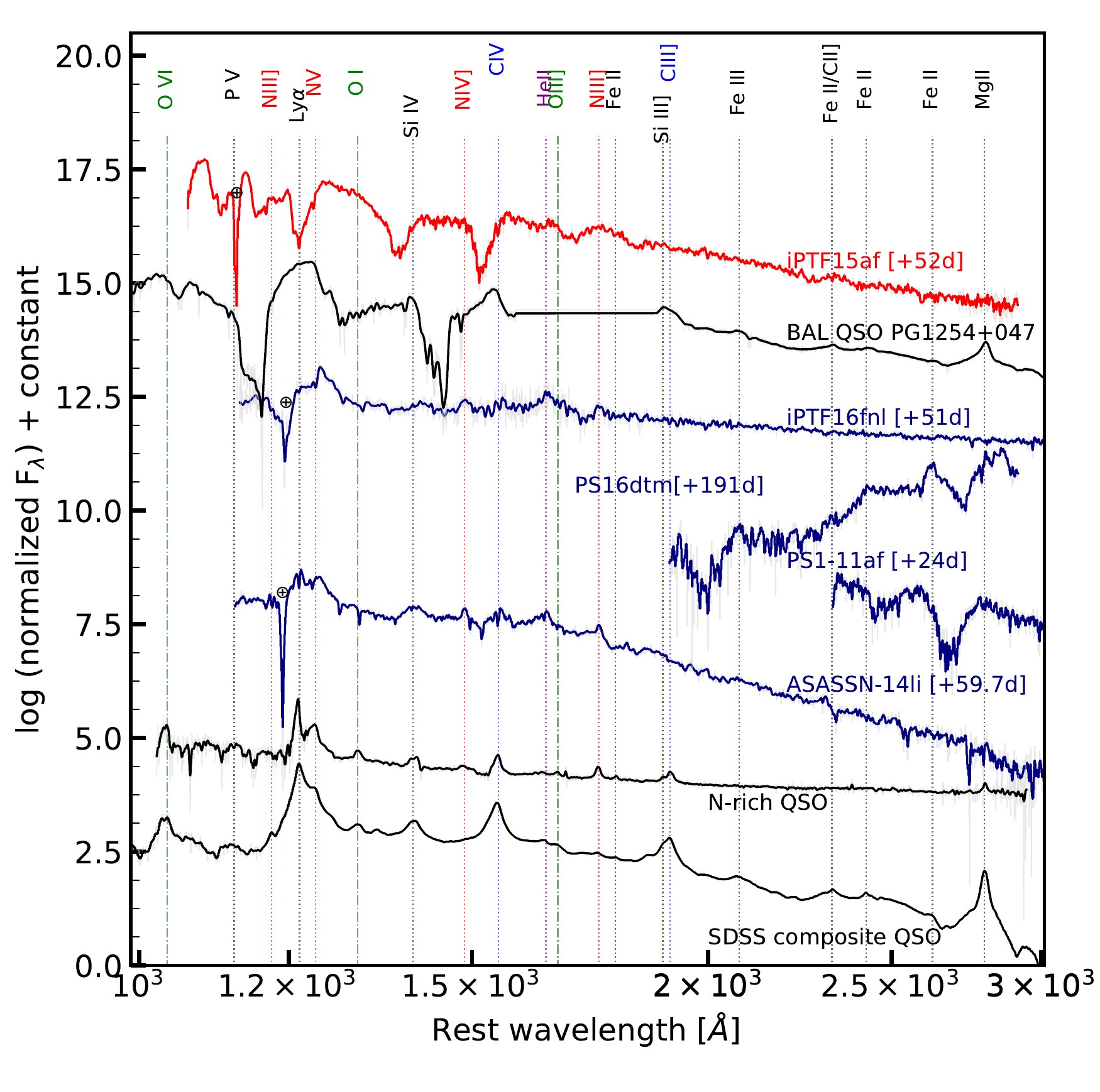} 
\caption{Comparison of {\it HST}/STIS spectrum of iPTF15af (red) with UV spectra of other TDEs (blue) and QSO (black). From top to bottom the comparison spectra are the BAL QSO PG1254+047 \cite{Hamann1998ApJ} (combined {\it HST} Faint Object Spectrograph (FOS) and SDSS spectrum), TDE iPTF16fnl \citep{Brown2018MNRAS}, TDE PS1-11af \citep{Chornock2014ApJ}, TDE PS16dtm \citep{Blanchard2017ApJ}, TDE ASASSN-14li \citep{Cenko2016ApJ}, N-rich QSO spectrum of SDSS J164148.19+223225.2 \citep{BatraBaldwin2014MNRAS}, and the SDSS composite QSO spectrum \citep{VandenBerk2001AJ}. The phase after discovery or peak for each TDE is shown in brackets. The spectra have been corrected for Galactic extinction. The symbol $\oplus$ shows the region with strong Galactic Ly$\alpha$ absorption. The most important lines identified in the spectra are marked at the top.}
\label{fig:spectra_uv_comparison}
\end{figure*}

iPTF15af is one of the four optical TDEs spectroscopically observed in the NUV and FUV. From these, PS16dtm and ASASSN-14li were discovered in galaxies with low-ionization nuclear emission-line regions. The latter was previously detected in X-ray and radio. Although the host of iPTF16fnl did not show any signs of activity, this event was the faintest and fastest TDE amongst all the optical sample explored so far. In this context, iPTF15af is quite representative of the general TDE population that we have seen so far: slowly evolving events with peak luminosities around $10^{44}$\,\ergs, generally in non-active galaxies. iPTF15af also had non-detections in radio and had weak (if any) soft X-ray emission. The characteristics observed for this transient may be relevant for the interpretation of the bulk of TDEs observed at UV wavelengths.  

A comparison of iPTF15af with other optically discovered TDEs is shown in Figure \ref{fig:spectra_uv_comparison}. Several AGN spectra are also included: a BAL QSO from \cite{Hamann1998ApJ}, a composite QSO from the SDSS spectroscopic survey  \citep{VandenBerk2001AJ}, and the nitrogen-rich QSO J164148.19+223225.2 \citep{BatraBaldwin2014MNRAS}. 

An initial remarkable difference is that none of the early-time emission in TDEs shows the presence of low-ionization elements such as \io{Mg}{ii}\,(15.0\,eV) and \io{Fe}{ii}\,(16.2\,eV), which are commonly observed in AGN spectra. The origin of these lines is believed to be in the partly neutral, low-ionization regions in optically thick gas clouds, which are being irradiated by X-ray photons from the central engine. For such lines to exist, the BLR clouds would require a lower limit on the column density on the order of $10^{22}$\,cm$^{-2}$ \citep{AGN1990}.  The lack of strong lines redward of 1900\,\AA\ can be attributed to insufficient shielding from high-energy photons, which keep the ionization continuum in the cloud well above their ionizaiton potential of $\sim 16$\,eV. 

The absorption components for low-ionization ions \io{Mg}{ii} and possibly \io{Fe}{ii} can be observed for two TDEs in our comparison sample: PS1-11af \citep{Chornock2014ApJ} and PS16dtm \citep{Blanchard2017ApJ}, which also differ from other TDEs in optical wavelengths. The absorption in PS1-11af was not present in the spectrum taken at $-5$\,days, suggesting that it formed in an outflow outside of the continuum region, analogous to our interpretation of iPTF15af.

The higher redshift of iPTF15af allows us to reach shorter wavelengths, where we putatively identify broad absorption corresponding to the  ion \io{P}{v} $\lambda\lambda$1118, 1128. This element is also observed in $\sim5$\% of BAL QSOs and its presence implies the existence of large column densities of $N_{\rm{H}} \approx 10^{22}-10^{23}$\,cm$^{-2}$ \citep{Capellupo2017MNRAS}. In most QSOs, the \io{P}{v} lines do not appear saturated, providing an impression of an extreme metal-to-hydrogen abundance ratio for the circum-nuclear gas \citep{Hamann1998ApJ}. However, if the absorbing outflows are concentrated in small optically thick cloudlets, they only would cover a fraction of the background emission source, supplying an extra continuum component.

As most BAL QSOs are weak in X-rays \citep{Brandt2000ApJ,Gallagher2001ApJ}, we can argue that the high column density outflows in iPTF15af would also act as an absorber for the soft X-rays, which would explain the early-time non-detections. The delayed reflection of high-energy emission in these cloudlets would possibly generate a highly enhanced \io{He}{ii}/\halpha ratio, as observed in several optical TDEs \citep{SaxtonPerets2018MNRAS}. However, the drop in density at late times would allow this emission to leak out, explaining the soft X-ray brightening of ASASSN-15oi reported by \cite{Gezari2017ApJ,Holoien2018MNRAS}. This lower density would also contribute to the narrowing of the optical line profiles, as observed in most optical TDEs \citep{Roth2018ApJ}. 

The lack of detectable radio and X-ray emission for iPTF15af further confirms that optical TDEs are predominantly radio quiet \citep{vanVelzen2013}, which seems to contrast with the high radio-loud fraction of AGNs among the nitrogen-rich population  \citep{Jiang2008ApJ}.

One traditional diagnostic of the gas density is the strength of the line \io{C}{iii}] $\lambda$1909, which has been extensively used in quasars to derive an upper limit for the density of the line-forming gas region \citep{Osterbrock1970ApJ}. Collisional de-excitation of this line becomes comparable to the spontaneous radiation rate at values approaching its critical density $n_c \approx 3\times 10^9$\,cm$^{-3}$. Hence, \io{C}{iii}] and other semiforbidden lines such as \io{N}{iii}] $\lambda$1750 ($n_c \approx 2 \times 10^{10}$\,cm$^{-3}$), \io{O}{iii}] $\lambda$1663 ($n_c \approx 4.6 \times 10^{10}$\,cm$^{-3}$), and \io{N}{iv}] $\lambda$1486 ($n_c \approx 3.4 \times 10^{10}$\,cm$^{-3}$), are expected to be weaker in the interior of the BLR, where the density exceeds their critical value. These lines are expected to gain importance in the external part of the BLR, where the density is lower. 

In the case of iPTF15af and other TDEs, we encounter a puzzle, as the spectrum shows unequal contributions for \io{C}{iii}] and \io{N}{iii}]. The nitrogen atom has similar ionization potential to carbon and the same ionization structure dependencies. Naively, we would expect the less denser BLR regions to form both lines at the same time, with the \io{N}{iii}]/\io{N}{iv}\ ratio enhanced in a similar way to \io{C}{iii}]/\io{C}{iv}. However, as previously discussed in the literature \citep{Cenko2016ApJ,Brown2018MNRAS}, this assumption fails for TDE spectra. While the semi-forbidden \io{N}{iii}] lines are clearly detected with broad velocity spread, the \io{C}{iii}] is lacking in all three TDEs having FUV spectra. The origin for this discrepancy has been associated with the nitrogen-enhanced composition of the stellar debris for super-solar-mass stars \citep{Kochanek2016MNRAS_abundancies,Yang2017ApJ}. The carbon-nitrogen-oxygen cycle would have substantially altered the star's composition with enhanced nitrogen at the expenses of carbon, translated into the observed line ratios. The debris of such a star would pollute the broad-line region (BLR), providing a possible link with nitrogen-rich QSOs \citep{Kochanek2016MNRAS_abundancies}.

Additionally, the existence of strong nitrogen emission around AGNs has been discussed by several recent studies. Some support the hypothesis that strong nitrogen lines imply highly super-solar metallicity in the vicinity of the SMBH \citep{BatraBaldwin2014MNRAS}, increased by continuous circumnuclear star formation. Others argue that the observed strong emission is most likely attributed to an exceptionally nitrogen-rich composition in the BLR, rather than overall enhanced metallicity \citep{Matsuoka2017AA}. The origin for this N-rich material would be linked to strong winds from the young asymptotic giant branch (AGB) stellar population in the nuclear regions of the galaxy.  The studies agree, though, that the N-loud AGN sample has on average lower SMBH masses as compared to the general population in the sample. This is a good indication that these galaxies are just initiating their main growth stage via high accretion episodes.

TDE discoveries have been predominantly associated with post-starburst galaxies having high stellar densities in the bulge \citep{French2016ApJ,Law-Smith2017ApJ,Graur2018ApJ}. For these galaxies, we would expect a young stellar population actively polluting the circumnuclear regions with both metal-rich supernova winds and N-enhanced AGB stellar winds. Regardless of the enrichment mechanism, TDEs would primarily be hosted in galaxies with already altered chemical abundances, translated into unusually strong nitrogen component in their UV spectra. However, an additional mechanism would need to be invoked to explain the apparent lack of \io{C}{iii}] and strong \io{C}{iv} emission, suggesting that the composition of the star indeed plays an important role in shaping the observed UV signature of TDEs.

One important difference between iPTF15af and BAL QSOs is the shape of the absorption. While the TDE shows a smooth gradient from lower to higher velocities and a sharp cut at $v_{\rm max}\approx9000$\,\kms, BAL QSOs generally exhibit the opposite trend: their absorption profile is detached from their emission. Their spectra show a sharp cut at lower velocities and a broader wing corresponding to higher velocity outflows. In BAL QSOs this means that the gas in our line of sight has already been radiatively accelerated to higher speeds. Closer to the emitting region, the gas will have higher column density, but lower velocity, appearing as a sharp trough in the spectra next to the emission line. In the case of iPTF15af we see that most of the absorption is caused by high-velocity gas transitioning to lower-velocity outflows. 

Outflows with $\sim10^4$\,\kms have been predicted to occur in TDEs with radiatively inefficient accretion flows \citep{MetzgerStone2016MNRAS}. Here we present UV spectroscopic data that confirms the existence of the outflow, and discuss the acceleration mechanism. For a given SMBH mass, the maximum velocity of the TDE outflow will depend on the Eddington ratio of the flare, the density of the gas and its distance to the central engine, as shown for AGN \citep{Risaliti2010}. At smaller radii, the gas will become overionized and the radiative wind will fail. At larger radii, the UV radiation field will be too weak to provide a noticeable acceleration. Therefore, the highest velocity for the bulk of stellar debris is likely to be achieved for dense gas at intermediate radius ($\sim$100\,R$_s$). The remaining bound material returning to the SMBH will progressively feel weaker acceleration, as the central source will be quickly fading with time, but also the material will be less dense. As the density of the ejecta decreases, it will become easier for the ionizing radiation to eventually penetrate this material
and escape to infinity, making the mechanism less efficient. Continuous monitoring of these absorption-line profiles can provide a new tool to better understand the outflow geometry in TDEs.

\section{Conclusions} \label{sec:conclusions}

iPTF15af is a TDE discovered in the core of a galaxy with signs of a short recent starburst episode and high nuclear stellar density in its core. The photometric optical and UV evolution of this event is consistent with previously studied TDEs, such as PS1-10jh and ASASSN-15oi. The event has a comparably slow rise time of $\sim60$\,days and peaks at $L_{\rm peak} \approx 1.5 \times 10^{44}$\,\ergs, with an estimated temperature of $T_{\rm{BB}}= (40-50) \times 10^{4}$\,K. 

The optical spectral evolution of iPTF15af shows broad characteristic \he2 and later \halpha lines superposed on a blue featureless continuum, which persists beyond four months past discovery. During the first three months the optical lines exhibits a fast evolution in their line profile, with the \halpha line appearing and becoming narrower at later times. The lines are no longer detected after $\sim120$\,days post-discovery.  

Our spectroscopic analysis reveals fluorescence lines of \io{O}{iii} and \io{N}{iii}, likely related to the Bowen fluorescence mechanism observed in planetary nebulae. The fluorescence between \io{He}{ii}$-$\io{O}{iii} and \io{O}{iii}$-$\io{N}{iii} would explain the pumping of extreme-UV flux into optical emission. 

We find that the medium producing the absorption in iPTF15af would come from a highly ionized source shielded by high-density material, justifying the lack of low-ionization lines of \io{Mg}{ii} and \io{Fe}{ii} and the appearance of broad absorption lines for highly ionized states of C, N, Si, and P.  The density of the gas would act as an absorber of the flare in X-rays, irradiating it toward lower-density gas enriched with nitrogen and/or metals supplied by a young stellar population in the bulge. The composition of the gas would then tip the N/C ratio to higher values than usually found in AGNs. In addition, the nitrogen enhanced and carbon depleted stellar debris of an evolved star can help justify the observed lack of \io{C}{iii} in the UV spectrum.

Contrary to BAL QSOs, the broad absorption line profiles observed in iPTF15af suggest that the highest column density material is moving at the outflow maximum velocity. We propose that the radiation pressure generated by the TDE flare at early times can supply the acceleration mechanism for this high-density gas. Future multi-epoch observations of TDEs in the UV would help to constrain the geometry, density and kinematics of the absorbing material, allowing to test this hypothesis. Late-time observations in X-rays are required to verify the change in the outflow's density, allowing high energy to escape.


\section{Acknowledgements} \label{sec:acknowledgements}

\begin{small}
We thank the anonymous referee, whose comments and suggestions helped to improve the paper. We are grateful to Peter Maksym, Sterl Phinney, Eliot Quataert, Cl\'ement Bonnerot, Tiara Hung, and Sjoert Van Velzen for valuable discussions and comments on this work. We are grateful to Eric Bellm, Melissa Graham, Anna Ho, and Daniel Perley for carrying out some of the spectroscopic observations and/or reductions, and Linda Strubbe for her support with the \textit{HST} proposal. We thank the \textit{HST} staff for the prompt scheduling of these ToO observations, as well as the PI Neil Gehrels and the \textit{Swift} ToO team for the timely execution of our observations.

This work was supported by the GROWTH project funded by the National Science Foundation (NSF) under grant AST-1545949. 
This work is part of the research programme VENI, with project number 016.192.277, which is (partly) financed by the Netherlands Organisation for Scientific Research (NWO).
Support for I.A. was provided by NASA through the Einstein Fellowship Program, grant PF6-170148. 

A.H. acknowledges support by the I-Core Program of the Planning and Budgeting Committee and the Israel Science Foundation. This research was supported by a Grant from the GIF, the German-Israeli Foundation for Scientific Research and Development.
A.V.F.'s group at UC Berkeley is grateful for financial assistance from the TABASGO Foundation, the Christopher R. Redlich Fund, and the Miller Institute for Basic Research in Science (UC Berkeley).

Some of the data presented herein were obtained at the W. M. Keck Observatory, which is operated as a scientific partnership among the California Institute of Technology, the University of California, and the National Aeronautics and Space Administration (NASA); the observatory was made possible by the generous financial support of the W. M. Keck Foundation.
This work makes use of observations from Las Cumbres Observatory. This publication makes use of data products from the Two Micron All Sky Survey, which is a joint project of the University of Massachusetts and the Infrared Processing and Analysis Center/California Institute of Technology, funded by NASA and the NSF. This publication makes use of data products from the Wide-field Infrared Survey Explorer, which is a joint project of the University of California, Los Angeles, and the Jet Propulsion Laboratory/California Institute of Technology, funded by NASA. Based on observations made with the NASA Galaxy Evolution Explorer. GALEX is operated for NASA by the California Institute of Technology under NASA contract NAS5-98034.  
\end{small}

\facilities{HST (STIS), Karl G. Jansky Very Large Array, Keck:I (LRIS), PO:1.2m, PO:1.5m, Swift,  WISE }
\software{ Astropy \citep{Astropy2018AJ}, emcee \citep{Foreman-Mackey2013PASP}, FPipe \citep{Fremling2016}, lcogtsnpipe \citep{Valenti16}, lmfit \citep{lmfit_newville_2014}, lpipe \url{http://www.astro.caltech.edu/~dperley/programs/lpipe.html}, MOSFIT  \citep{Guillochon2018ApJS},  NumPy \citep{NumPy2011}, PTFIDE \citep{Masci2017PASP}, SciPy \citep{SciPy}}

\small
\bibliographystyle{yahapj}
\bibliography{15af}


\end{document}